\documentclass[11pt,draftclsnofoot,final,onecolumn,notitlepage]{IEEEtran}

\IEEEoverridecommandlockouts

\usepackage{cite}
\usepackage{ieeefig}
\usepackage{amsfonts}
\usepackage{epsfig,color}
\usepackage{graphicx}
\usepackage[caption=false,font=footnotesize]{subfig}
\usepackage{stfloats}
\usepackage{amsmath}
\usepackage{paralist}
\usepackage{color}

\newtheorem{theorem}{\bf Theorem}
\newtheorem{lemma}{\bf Lemma}
\newtheorem{corollary}{\bf Corollary}
\newtheorem{proposition}{\bf Proposition}
\newtheorem{remark}{\bf Remark}
\newtheorem{definition}{\bf Definition}
\newtheorem{design}{\bf Design Criterion}
\newcommand{\define}{\stackrel{\triangle}{=}}

\begin{document}
\title{\LARGE{The Degrees of Freedom Region of the MIMO Interference Channel with Shannon Feedback} }

\author{Chinmay S.~Vaze %
        and~Mahesh K.~Varanasi% <-this % stops a space
\thanks{This work was supported in part by NSF EAGER Grant CCF-1144026. %
The authors are with the Department of Electrical, Computer, and Energy
Engineering, University of Colorado, Boulder, CO 80309-0425 USA
(e-mail: {Chinmay.Vaze, varanasi}@colorado.edu).
The material in this paper was presented in part
at the $49^{th}$ Annual Allerton Conference on Communication, Control, and Computing, Monticello, IL, USA, Sep. 28-30, 2011.} }

%\markboth{Journal of \LaTeX\ Class Files,~Vol.~6, No.~1, January~2007}%
%{Shell \MakeLowercase{\textit{et al.}}: Bare Demo of IEEEtran.cls for Journals}

% \markboth{}{C. S. Vaze and M. K. Varanasi: Short Title}


\maketitle

\begin{abstract}
The two-user multiple-input multiple-output (MIMO) fast-fading interference channel (IC) with an arbitrary number of antennas at each of the four terminals is studied under the settings of Shannon feedback, limited Shannon feedback, and output feedback, wherein all or certain channel matrices and outputs, or just the channel outputs, respectively, are available to the transmitters with a finite delay. While for most numbers of antennas at the four terminals, it is shown that the DoF regions with Shannon feedback and for the limited Shannon feedback settings considered here are identical, and equal to the DoF region with just delayed channel state information (CSIT), it is shown that this is not always the case. For a specific class of MIMO ICs characterized by a certain relationship between the numbers of antennas at the four nodes, the DoF regions with Shannon and the limited Shannon feedback settings, while again being identical, are strictly bigger than the DoF region with just delayed CSIT. To realize these DoF gains with Shannon or limited Shannon feedback, a new retrospective interference alignment scheme is developed wherein transmitter cooperation made possible by output feedback in addition to delayed CSIT is employed to effect a more efficient form of interference alignment than is feasible with previously known schemes that use just delayed CSIT. The DoF region for just output feedback, in which each transmitter has delayed knowledge of only the receivers' outputs, is also obtained for all but a class of MIMO ICs that satisfy one of two inequalities involving the numbers of antennas.% at each node. %The same outer bound on the DoF region for all scenarios for which DoF regions are obtained is also shown to hold under a more general form of feedback than Shannon feedback called designable Shannon feedback in which both transmitters have all delayed CSIT and some designable functions of delayed channel states and outputs from both receivers.
\end{abstract}


% \IEEEpeerreviewmaketitle

\newpage
% %%%%%%%%%%%%%%%%%%%%%% DEFINE FOLLOWING ACRONYMS IN THE INTRODUCTION %%%%%%%%%%%
% IC
% DoF
% CSI, CSIR, CSIT
% MIMO
% \mathcal{C}\mathcal{N}(0,I_{N_i})$ everything in it.
% RV
% i.i.d.
% IA
% BC

\section{Introduction}

\IEEEPARstart{T}{he} characterization of the capacity of channels with feedback, where the channel outputs are known to the transmitter(s) with a finite delay, is a classical problem in information theory. For example, it is well known that feedback can not increase the capacity of a memoryless point-to-point channel \cite{CT}. Moreover, feedback can not increase the capacity of a point-to-point channel with additive, correlated Gaussian noise by more than one bit \cite{CT}. Interestingly, multi-user channels exhibit a different behavior. In particular, feedback can enhance the capacity of even the memoryless multiple access channel (MAC) \cite{gaarder_wolf_DM_MAC_fb_1975, ozarow_fb_capacity_mac_1984} but again this improvement is bounded for the Gaussian MAC \cite{ozarow_fb_capacity_mac_1984}. There has also been much interest in characterizing the capacity region of other memoryless networks with feedback such as the broadcast channel (BC) \cite{Gamal_fb_capacity_degraded_BC}. However, due to the apparent intractability of such problems for more complex topologies, capacity {\em approximations} have been sought. Of these approximate capacity metrics, the degrees of freedom (DoF) region has received considerable attention. The DoF region denotes the rate of growth of the capacity region with respect to the logarithm of the signal-to-noise ratio (SNR) in the limit of asymptotically high SNR. For example, it can be deduced from  \cite{Chiachi-Jafar} that for the $2$-user Gaussian MIMO IC  output feedback can not enhance the DoF region when there is perfect and instantaneous CSIT. In \cite{suh_tse_feedback_capacity_SISO_IC_2bits}, the feedback capacity region is characterized to within a constant gap of $2$ bits for the single-antenna (or SISO) IC. %They also proved that feedback can enhance the so-called {\em generalized} DoF region, i.e., the DoF region may expand only if the channel coefficients are assumed to vary with the SNR in a specific manner. %However, this setting is only of a theoretical interest since the channel coefficients can not change with the SNR in practice.
Further, it is well known that feedback fails to improve the DoF regions of the Gaussian MIMO MAC and the Gaussian MIMO BC. %and the $K$-user Gaussian MIMO IC with an equal number of antennas at all terminals.
It is not yet known if there are networks with instantaneous CSIT for which (output) feedback enhances the DoF region.

%The importance of having feedback in the no-CSIT setting can be understood on noting that the complete lack of CSIT severely constrains the design of transmit signals and thus results in a relatively high amount of inter-user interference at the receivers, which can be mitigated to a large extent by the transmitters using feedback to achieve a bigger DoF region.

While the DoF of networks under the idealized assumptions of perfect, often global, and instantaneous CSIT have been well studied for numerous networks, the much more conservative setting of isotropic fading with transmitters having no CSIT has recently been extensively studied as well \cite{Jafar-Goldsmith, Lapidoth, Chiachi2, Vaze_Dof_final, Zhu_Guo_noCSIT_DoF_2010, Vaze_Dof_Cognitive_IC_ISIT, Vaze_Varanasi_Interf_Loclzn_2011, GDoF_region_noCSIT_IC_isit11}. Networks without CSIT but with (output) feedback have also been considered from which it is known that in the absence of CSIT feedback can enhance the DoF regions of the $K$-user BC \cite{maddah_ali_tse_delayed_CSIT, Vaze-Varanasi-delay-MIMOBC, abdoli_3user_BC_delayed_isit}, the $2$-user SISO X channel \cite{Jafar_Shamai_retrospective_IA}, and $3$-user SISO IC \cite{Jafar_Shamai_retrospective_IA,Vaze_Dof_final}. Thus, unlike the instantaneous CSIT case, feedback can be beneficial even from the DoF perspective when there is no CSIT. This suggests that the benefit of feedback depends critically on the availability of CSIT since it is vastly different at the two extremes of having instantaneous CSIT and having no CSIT whatsoever. %With this backdrop, we analyze conditions that lie intermediate between the two extremes.

Moving beyond models that are either too idealized on the one extreme, or too conservative on the other, we consider here the delayed CSIT model wherein the channel state varies independently across time and the transmitters know perfectly the {\it past} channel states (cf. \cite{maddah_ali_tse_delayed_CSIT,Jafar_Shamai_retrospective_IA,Vaze-Varanasi-delay-MIMOBC}).  For such a setting we investigate the question of whether output feedback can improve the DoF region. Clearly, this question can be definitively answered only for networks for which the DoF with (just) delayed CSIT are known, of which there are but few. Of all networks for which the DoF are known except for the MIMO IC, this question has so far been answered in the negative.  In particular, it is known that with delayed CSIT output feedback can not increase (a) the sum-DoF of the $K$-user MISO BC with at least $K$ transmit antennas \cite{maddah_ali_tse_delayed_CSIT}, (b) the DoF region of the $2$-user MIMO BC \cite{Vaze-Varanasi-delay-MIMOBC}, (c) the sum-DoF of the $3$-user MIMO BC with $N$ antennas at all three receivers and at most $2N$ antennas at the transmitter \cite{abdoli_3user_BC_delayed_isit} and (d) the $2 \times 2 \times 2$ interference network \cite{vaze_varanasi_2x2x2_allerton11}.

\begin{figure}[t]
\begin{centering}
\includegraphics[bb=0bp 50bp 720bp 510bp,clip,scale=0.375]{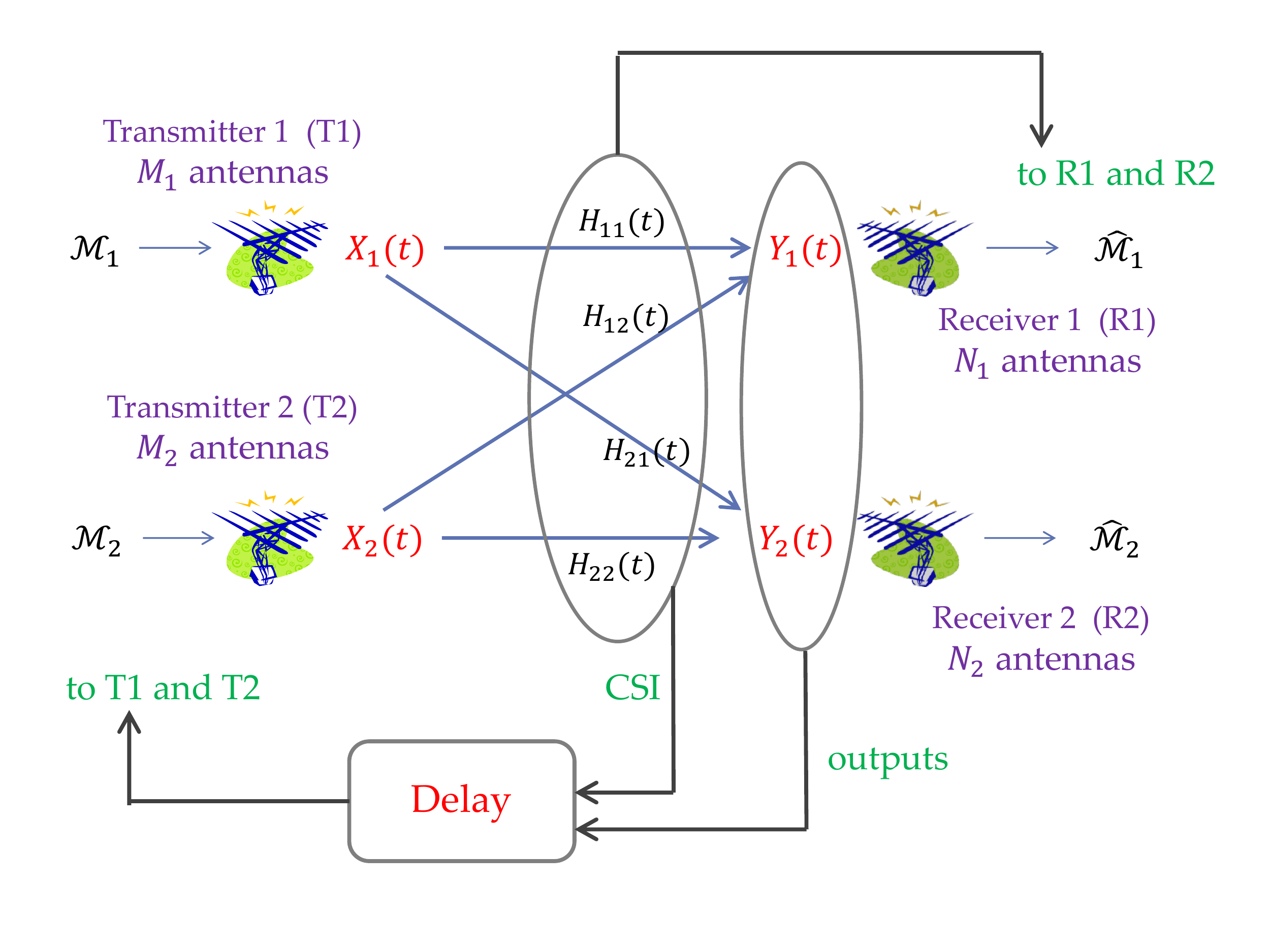}
\par\end{centering}
\caption{The $(M_1,M_2,N_1,N_2)$ MIMO IC with Shannon Feedback} \label{fig: IC schematic dCSI+opfb 2user IC}
\end{figure}

The only other exact characterization for the DoF region with (just) delayed CSIT is provided by the authors in \cite{Vaze_Varanasi_delay_MIMO_IC} for the two-user MIMO IC with an arbitrary number of antennas at each of the four terminals. Consequently, by obtaining the complete DoF region of this general two-user MIMO IC with delayed CSIT and output feedback (i.e., under Shannon feedback), and showing that for some cases there is a strict enhancement of the DoF region over that with just delayed CSIT, we answer the question of whether output feedback can enhance the DoF region of a network with delayed CSIT for the first time in the affirmative. %In other words, all other relevant results \cite{Jafar_Shamai_retrospective_IA, Ghasemi_Khandani_X_delayed, abdoli_3user_BC_delayed_isit, maddah_ali_tse_delayed_CSIT} are in the form of the lower-bounds on the delayed-CSIT DoF of different wireless networks.

%by the lack of rich literature on this topic, i.e., very few results are available, which derive the exact characterization of the DoF (region) of wireless networks with delayed CSIT. More precisely, apart from the four results, namely, \cite{maddah_ali_tse_delayed_CSIT, abdoli_3user_BC_delayed_isit, Vaze-Varanasi-delay-MIMOBC, vaze_varanasi_2x2x2_allerton11} cited earlier, the only other known exact characterization is provided by \cite{Vaze_Varanasi_delay_MIMO_IC}, which determines the delayed-CSIT DoF region of the $2$-user MIMO IC with arbitrary number of antennas at four terminals. In other words, all other results are in the form of the lower-bounds on the delayed-CSIT sum-DoF of different wireless networks \cite{Jafar_Shamai_retrospective_IA, Ghasemi_Khandani_X_delayed, abdoli_3user_BC_delayed_isit, maddah_ali_tse_delayed_CSIT}.

In particular, it is shown here that if $M_i$ and $N_i$, respectively, denote the number of antennas at the $i^{th}$ transmitter and the $i^{th}$ receiver of a two-user MIMO IC, then the DoF region with Shannon feedback is strictly bigger than the corresponding delayed-CSIT DoF region, if and only if one of the two inequalities, namely,
\[
M_1 > N_1+N_2-M_2 > N_1 > N_2 > M_2 > N_2 \frac{N_2-M_2}{N_1-M_2}
\]
or its symmetric counterpart (obtained by switching the user indices), holds. For MIMO ICs for which neither of these two inequalities holds output feedback does not improve the delayed CSIT DoF region. %as with all previously known results for the MIMO BC and the $2 \times 2 \times 2$ interference network.

To derive our main result, we first obtain an outer-bound to the DoF region with Shannon feedback. This outer bound is then shown to be achievable for all but the above described class of MIMO ICs using just delayed CSIT. For the class where the DoF region is strictly larger than that with delayed CSIT, we develop a new retrospective interference alignment scheme in which each transmitter -- using the side information available to it -- reconstructs and transmits the previously transmitted signal of the other transmitter to provide an opportunity to its paired receiver to cancel the interference it encountered at a previous time instant, while simultaneously delivering new useful linear combinations to the unpaired receiver. Consequently, Shannon feedback induces a new form of transmitter cooperation which is key to realizing the DoF gains attainable with Shannon feedback over that with just delayed CSIT.

Moreover, it is seen that to achieve this more efficient interference alignment all of the channel matrices and outputs are not needed at both transmitters. In particular, two limited Shannon feedback settings are described that are sufficient to achieve the DoF region with Shannon feedback. It is also observed that if in addition to delayed CSIT the feedback is allowed to be some designable function of past channel outputs and states, a setting that is more optimistic than Shannon feedback, the DoF region doesn't expand over that of the DoF region in the Shannon feedback case. Furthermore, with just output feedback without any form of CSIT, it is shown that the DoF region is the same as that for delayed CSIT with the exception of a class of MIMO ICs characterized by one of two inequalities involving the numbers of antennas at the four terminals. For this class, the DoF region remains an open problem at this time.

The rest of the paper is organized as follows. Section \ref{subsec: channel model Shannon journal} describes the model of MIMO IC under various assumptions about feedback, states the main results of this work on the DoF regions under these assumptions and provides an example of the new retrospective interference alignment scheme. Proofs of the results are contained in Sections \ref{sec: proof of thm: outer-bound 2user IC d-CSI and op fb} and \ref{sec: proof of thm: inner-bound 2user IC d-CSI and op fb} and the appendix.

\section{Channel Models, Main Results, and IA with Shannon Feedback}

In Section \ref{subsec: channel model Shannon journal}, the MIMO IC model with Shannon feedback, limited Shannon feedback, output feedback and designable Shannon feedback are described. Section \ref{subsec: main results dCSI_op 2user IC} contains the main results on the DoF regions under these settings. In Section \ref{subsec: IA example dCSI_opfb 2user IC}, we discuss how interference alignment can be achieved with Shannon and limited Shannon feedback. Section \ref{subsec: intuition behind results dCSI_opfb 2user IC} provides some insight on the main results.

\subsection{The MIMO IC with Shannon Feedback}  \label{subsec: channel model Shannon journal}

The MIMO IC consists of two transmitters, denoted as T1 and T2, and their corresponding receiver, labeled R1 and R2, respectively, as in Fig. \ref{fig: IC schematic dCSI+opfb 2user IC}. The $(M_1,M_2,N_1,N_2)$ MIMO IC is defined via the input-output relationships
\begin{eqnarray}
Y_1(t) & = & H_{11}(t) X_1(t) + H_{12}(t) X_2(t) + W_1(t), \\
Y_2(t) & = & H_{21}(t) X_1(t) + H_{22}(t) X_2(t) + W_2(t),
\end{eqnarray}
where, at time $t$, $Y_i(t) \in \mathbb{C}^{N_i \times 1}$ is the signal received by the $i^{th}$ receiver; $X_i(t) \in \mathbb{C}^{M_i \times 1}$ is the signal transmitted by the $i^{th}$ transmitter; $H_{ij}(t) \in \mathbb{C}^{N_i \times M_j}$ is the channel matrix between the $i^{th}$ receiver and the $j^{th}$ transmitter; $W_i(t) \in \mathbb{C}^{N_i \times 1}$ is the additive white Gaussian noise at the $i^{th}$ receiver; and there is a power constraint of $P$ on the transmit signals, i.e., $\mathbb{E} ||X_i(t)||^2 \leq P$ $\forall$ $i,t$.

\begin{figure}[t]
\begin{centering}
\includegraphics[bb=30bp 100bp 660bp 480bp,clip,scale=0.4]{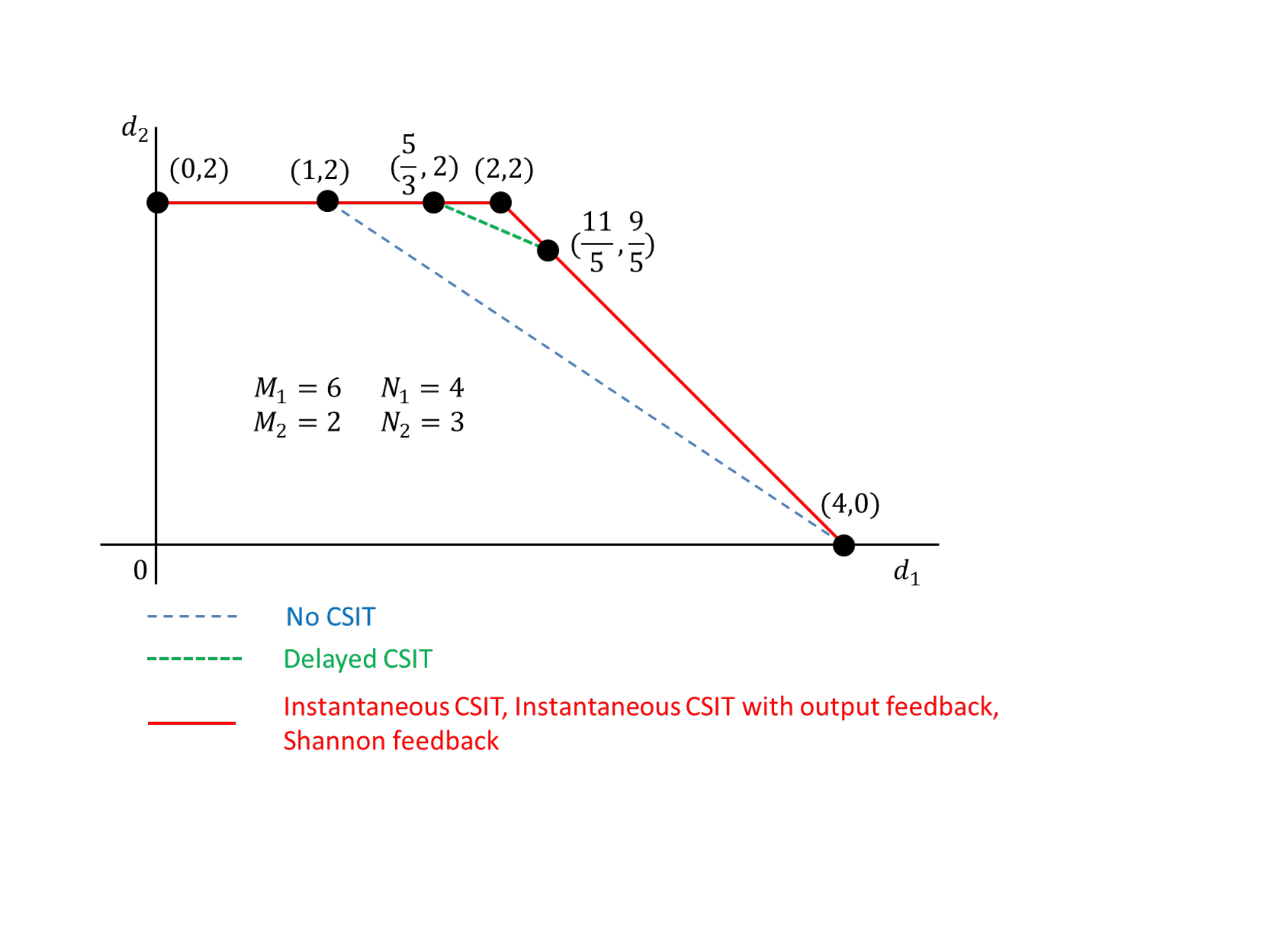}
\par\end{centering}
\caption{Comparison of the DoF regions of a MIMO IC} \label{fig: DoF region comparison dCSI+opfb 2user IC}
\end{figure}

For simplicity, we study here the case of additive white Gaussian noise, i.e., all entries of $W_i(t)$, $i=1,2$, are independent and identically distributed (i.i.d.) according the complex normal distribution with zero-mean and unit-variance (denoted, henceforth, as $\mathcal{C}\mathcal{N}(0,1)$). Further, we assume that the channel matrices are i.i.d. Rayleigh faded, i.e., all elements of all channel matrices are i.i.d. according to $\mathcal{C}\mathcal{N}(0,1)$ distribution (denoted as i.i.d. $\sim \mathcal{C}\mathcal{N}(0,1)$). Next, it is assumed that the realizations of additive noises and channel matrices are i.i.d. across time and that they are mutually independent of each other.

Throughout this paper, both receivers are taken to know all channel matrices perfectly. Since there is no delay constraint on decoding it is assumed, without loss of generality, that CSI is available instantaneously to the receivers. Further, all terminals are always assumed to know the distribution of the channel matrices.

We start by defining the term Shannon feedback (cf. \cite{Jafar_Shamai_retrospective_IA}) and later consider other types of feedback. Here, the two transmitters are assumed to know the channel matrices and the channel outputs perfectly with a finite delay. This delay is taken to be of $1$ symbol time without loss of generality. In particular, the channel matrices $\{H_{ij}(t)\}_{i,j=1}^2$ and the channel outputs $Y_1(t)$ and $Y_2(t)$ are taken to be known perfectly to both transmitters at time $t+1$. %We next introduce some notation.
\newline
{\em Notation: } For $n \geq 0$, $\overline{H}(n) \define \big\{ H_{11}(t), H_{12}(t), H_{21}(t), H_{22}(t) \big\}_{t=1}^n$ if $n \geq 1$ and $\overline{H}(n) = \phi$ ($\phi=$ some constant) if $n = 0$. Similarly, for each $i \in \{1,2\}$, $\overline{Y}_i(n) = \{Y_i(t)\}_{i=1}^n$ if $n \geq 1$ and $\overline{Y}_i(n) = \phi$ if $n = 0$.

Let $\mathcal{M}_1$ and $\mathcal{M}_2$ be two independent messages to be sent by T1 and T2, respectively, over a block length of $n$, where the message $\mathcal{M}_i$ is intended for the $i^{th}$ receiver. It is assumed that $\mathcal{M}_i$ is distributed uniformly over a set of cardinality $2^{nR_i(P)}$, $R_i(P) \geq 0$, when there is a power constraint of $P$ at the transmitters. A coding scheme for blocklength $n$ consists of two encoding functions $f_i^{(n)} = \{f_{i,t}^{(n)}\}_{t=1}^n$, $i = 1,2$, such that
\[
X_i(t) = f_{i,t}^{(n)} \Big( \mathcal{M}_i, \overline{H}(t-1), \overline{Y}_1(t-1), \overline{Y}_2(t-1) \Big) \mbox{ with } \mathbb{E} ||X_i(t)||^2 \leq P ~ \forall ~i,t,
\]
and two decoding functions such that
\[
\hat{\mathcal{M}}_i = g_i^{(n)} \Big( \overline{Y}_i(n), \overline{H}(n) \Big)  \mbox{ where } i \in \{1,2\}.
\]
A rate tuple $\big( R_1(P),R_2(P) \big)$ is said to be achievable if there exists a sequence (over $n$) of coding schemes such that probability of $\mathcal{M}_1 \not= \hat{\mathcal{M}}_1$ or $\mathcal{M}_2 \not= \hat{\mathcal{M}}_2$ tends to zero as $n \to \infty$.

The capacity region $\mathcal{C}(P)$ is defined to be the set of all achievable rate tuples $\big( R_1(P),R_2(P) \big)$ when the power constraint at the transmitters is $P$. The DoF region with Shannon feedback is defined as
\begin{eqnarray*}
\lefteqn{ \mathbf{D}^{\rm S} \define \Biggl\{ (d_1,d_2) \in \mathbb{R}^2_+ \Bigl| ~ \forall ~ (w_1,w_2) \in \mathbb{R}^2_+, \Biggr. } \\
&& {} \Biggl. ~ ~ w_1 d_1 + w_2 d_2 \leq \limsup_{P \to \infty} \frac{1}{\log_2 P} \Biggl[ \sup_{\big( R_1(P),R_2(P) \big) \in \mathcal{C}(P)} \Big\{ w_1 R_1(P) + w_2 R_2(P) \Big\} \Biggr] \Biggr\},
\end{eqnarray*}
where $\mathbb{R}^2_+$ denotes the set of pairs of non-negatives real numbers, and $\limsup_{P \to \infty}$ stands for the limit superior \cite{Royden} as $P \to \infty$. It can be easily proved that the DoF region $ \mathbf{D}^{\rm S}$ is closed \cite{Royden} and convex \cite{CT}.

Consider next the MIMO IC under four other settings defined below.
\begin{itemize}
\item {\em Designable Shannon feedback:} both transmitters know the channel matrices $\{H_{ij}(t)\}_{i,j}$ and {\em modified} channel outputs $\tilde{Y}_1(t)$ and $\tilde{Y}_2(t)$ at time $t+1$ (in general, with some finite delay), where, for each $i \in \{1,2\}$, $\tilde{Y}_i(t) \in \mathbb{C}^{N_i \times 1}$ is a deterministic function of $\overline{Y}_i(t)$ and  $\overline{H}(t)$.
\item {\em Limited Shannon feedback of Type 1}: each transmitter knows the other receiver's incoming channels and outputs with some delay (which, without loss of generality we take to be 1 time unit), i.e., the $i^{th}$ transmitter knows the channel matrices $H_{ji}(t)$ and $H_{jj}(t)$ and the received signal $Y_j(t)$ all at time $t+1$, for each $(i,j) \in \{(1,2), (2,1)\}$.
\item {\em Limited Shannon feedback of Type 2}: each transmitter is provided at each time its own receiver's outputs as well as the four channel matrices, all with some delay (which, without loss of generality we take to be 1 time unit); i.e., the $i^{th}$ transmitter knows $Y_i(t)$ and $\{H_{jk}(t)\}_{j,k=1}^2$ at time $t+1$.
\item {\em Output feedback:} both transmitters know the channel outputs $Y_1(t)$ and $Y_2(t)$ at time $t+1$ (or, in general, with a delay of finite number of time slots) but they have no knowledge of channel matrices whatsoever.
\end{itemize}
The DoF regions of the MIMO IC under these settings can be defined in analogous manner to that with Shannon feedback, and are denoted, respectively, as $\mathbf{D}^{\rm \mathit{d}S}$, $\mathbf{D}^{\rm \mathit{l}S1}$,  $\mathbf{D}^{\rm \mathit{l}S2}$, and $\mathbf{D}^{\rm op}$. Since the designable Shannon feedback setting is stronger than that of Shannon feedback and the limited Shannon feedback and output feedback settings are weaker, we have that
\[
\mathbf{D}^{\mathrm{\mathit{d}S}} \supseteq \mathbf{D}^{\mathrm{S}} \supseteq  \mathbf{D}^{\mathrm{\mathit{l}S1}}, \qquad \mathbf{D}^{\mathrm{S}}  \supseteq \mathbf{D}^{\mathrm{\mathit{l}S2}}, \qquad \mathbf{D}^{\mathrm{S}} \supseteq \mathbf{D}^{\mathrm{op}}.
\]

Furthermore, the conditions of delayed CSIT, instantaneous CSIT, and instantaneous CSIT with output feedback are defined as follows:
\begin{itemize}
\item {\em delayed CSIT:} the channel matrices $\{H_{ij}(t)\}_{i,j=1}^2$ are known to the transmitters at time $t+1$;
\item {\em instantaneous CSIT:} the channel matrices $\{H_{ij}(t)\}_{i,j=1}^2$ are known to the transmitters instantaneously (i.e., at time $t$); and
\item {\em instantaneous CSIT and output feedback:} the channel matrices $\{H_{ij}(t)\}_{i,j=1}^2$ are known to the transmitters at time $t$, and additionally, they know the channel outputs $Y_1(t)$ and $Y_2(t)$ at time $t+1$.
\end{itemize}
The DoF regions corresponding to these three assumptions can again be defined analogously, and are denoted, respectively, as $\mathbf{D}^{\mathrm{dCSI}}$, $\mathbf{D}^{\mathrm{iCSI}}$, and $\mathbf{D}^{\mathrm{iCSI \& op}}$. Clearly,
\[
\mathbf{D}^{\mathrm{dCSI}} \subseteq \mathbf{D}^{\mathrm{iCSI}}, \qquad \mathbf{D}^{\mathrm{S}} \subseteq \mathbf{D}^{\mathrm{iCSI \& op}}.
\]

\subsection{Main Results} \label{subsec: main results dCSI_op 2user IC}

The characterization of $\mathbf{D}^{\mathrm{iCSI \& op}}$ below asserts that, in the presence of instantaneous CSIT, output feedback can not improve the DoF region.
\begin{lemma}\label{lem: DoF region 2user IC p-CSI and op-fb}
For the MIMO IC with i.i.d. Rayleigh fading, the DoF region with instantaneous CSIT and output feedback is given by
\begin{eqnarray*}
\lefteqn{ \mathbf{D}^{\mathrm{iCSI \& op}} = \Big\{ (d_1,d_2) \Big| 0 \leq d_1 \leq \min(M_1,N_1), \; 0 \leq d_2 \leq \min(M_2,N_2) \Big. \Big. } \\
&& {} \qquad \qquad \qquad \Big. d_1 + d_2 \leq \min \big[ M_1+M_2, N_1+N_2, \max(M_1,N_2), \max(M_2,N_1) \big] \Big\} .
\end{eqnarray*}
Moreover, $\mathbf{D}^{\mathrm{iCSI}} = \mathbf{D}^{\mathrm{iCSI \& op}}$.
\end{lemma}
\begin{IEEEproof}
See Appendix \ref{app: proof of lem: DoF region 2user IC p-CSI and op-fb}.
\end{IEEEproof}
%This lemma yields the following simple corollary.
%\begin{corollary} \label{cor: DoF region perfectCSIT+opfb dCSI+opfb 2user IC}
%For the MIMO IC with i.i.d. Rayleigh fading, $\mathbf{D}^{\mathrm{iCSI}} = \mathbf{D}^{\mathrm{iCSI \& opfb}}$.
%\end{corollary}
%\begin{IEEEproof}
%Clearly, $\mathbf{D}^{\mathrm{iCSI}} \subseteq \mathbf{D}^{\mathrm{iCSI \& opfb}}$. Hence, it is sufficient to prove that the region $\mathbf{D}^{\mathrm{iCSI \& opfb}}$ is achievable with just instantaneous CSIT, which is known \cite[Theorem 4]{Chiachi-Jafar}, \cite{Jafar-Maralle}.
%\end{IEEEproof}
\vskip 1em

%Consider the following definition.
\begin{definition}
The region $\mathbf{D}^{\mathrm{S}}_{\mathrm{outer}}$ is defined as
\begin{eqnarray*}
\lefteqn{ \mathbf{D}^{\mathrm{S}}_{\mathrm{outer}} \define \Big\{ (d_1,d_2) \Big| ~ L_{0i} ~ \define ~ 0 \leq d_i \leq \min (M_i,N_i), ~ i=1,2; \Big. \Big. } \\
&& {} L_1 ~ \define ~ \frac{d_1}{\min(N_1 + N_2, M_1)} + \frac{d_2}{\min(N_2,M_1)} \leq \frac{\min (N_2,M_1+M_2)}{\min(N_2,M_1)}; \\
&& {} L_2 ~ \define ~ \frac{d_1}{\min(N_1,M_2)} + \frac{d_2}{\min(N_1 + N_2,M_2)} \leq \frac{\min(N_1, M_1+M_2)}{\min(N_1,M_2)}; \\
&& {} \left. L_3 ~ \define ~ d_1 + d_2 \leq \min \big[ M_1 + M_2, N_1+N_2, \max(M_1,N_2), \max(M_2,N_1) \big] \right\}.
\end{eqnarray*}
Note that the first two bounds on $d_1$ and $d_2$ have been denoted as $L_{01}$ and $L_{02}$ respectively; while the last three bounds on the weighted sums of $d_1$ and $d_2$ are denoted respectively by $L_1$, $L_2$, and $L_3$.
\end{definition}

%The following theorem shows that the region $\mathbf{D}^{\mathrm{S}}_{\mathrm{outer}}$ is an outer bound to $\mathbf{D}^{\mathrm{S}}$.
\vskip 1em
\begin{theorem}[Outer-Bound with Shannon feedback] \label{thm: outer-bound 2user IC d-CSI and op fb}
For the MIMO IC with i.i.d. Rayleigh fading, the DoF region with Shannon feedback is outer-bounded by the region $\mathbf{D}^{\mathrm{S}}_{\mathrm{outer}}$, i.e.,
\[
\mathbf{D}^{\mathrm{S}} \subseteq \mathbf{D}^{\mathrm{S}}_{\mathrm{outer}}.
\]
\end{theorem}
\begin{IEEEproof}
See Section \ref{sec: proof of thm: outer-bound 2user IC d-CSI and op fb}.
\end{IEEEproof}
\vskip 1em

%The following definition is useful in what follows.
\begin{definition}
For a given $i \in \{1,2\}$, Condition $i$ is said to hold whenever the inequality
\[
M_i > N_1 + N_2 - M_j > N_i > N_j > M_j > N_j \frac{N_j - M_j}{N_i - M_j}
\]
holds for $j \in \{1,2\}$ with $j \not= i$.
\end{definition}
Clearly, the two conditions are symmetric counterparts of each other. Moreover, the two conditions can not be true simultaneously. Condition $i$ can not hold if $N_j \geq N_i$.

%The next theorem states that the outer-bound of Theorem \ref{thm: outer-bound 2user IC d-CSI and op fb} is tight.
\vskip 1em
\begin{theorem}[The DoF Region with Shannon feedback] \label{thm: inner-bound 2user IC d-CSI and op fb}
For the MIMO IC with i.i.d. Rayleigh fading, the DoF region with Shannon feedback is equal to the region $\mathbf{D}^{\mathrm{S}}_{\mathrm{outer}}$, i.e.,
\[
\mathbf{D}^{\mathrm{S}} = \mathbf{D}^{\mathrm{S}}_{\mathrm{outer}}.
\]
\end{theorem}
\begin{IEEEproof}
It is sufficient to prove that the region $\mathbf{D}^{\mathrm{S}}_{\mathrm{outer}}$ is achievable when there is Shannon feedback. We assume without loss of generality that $N_1 \geq N_2$ (note, under this assumption, that Condition $2$ can not hold).

Suppose Condition $1$ does not hold. Then from \cite[Theorem 2]{Vaze_Varanasi_delay_MIMO_IC}, we observe that
\[
\mathbf{D}^{\mathrm{S}}_{\mathrm{outer}} = \mathbf{D}^{\mathrm{dCSI}},
\]
so that the theorem follows by noting that $\mathbf{D}^{\mathrm{dCSI}} \subseteq \mathbf{D}^{\mathrm{S}} \subseteq \mathbf{D}^{\mathrm{S}}_{\mathrm{outer}}$.

Thus, it is only required to show that the region $\mathbf{D}^{\mathrm{S}}_{\mathrm{outer}}$ is achievable when Condition $1$ holds. The detailed proof is given in Section \ref{sec: proof of thm: inner-bound 2user IC d-CSI and op fb}.
\end{IEEEproof}
\vskip 1em

The basic idea behind the  interference alignment (IA) based achievability scheme developed in Section \ref{sec: proof of thm: inner-bound 2user IC d-CSI and op fb} to prove the above theorem is illustrated via an example in Section \ref{subsec: IA example dCSI_opfb 2user IC} which shows that  $\mathbf{D}^{\mathrm{S}} \not= \mathbf{D}^{\mathrm{dCSI}}$ and provides insight as to why the DoF regions are not always identical. Further, Section \ref{subsec: intuition behind results dCSI_opfb 2user IC} compares the techniques used to achieve IA under Shannon feedback and under delayed CSIT.

\begin{remark}[Comparison of $\mathbf{D}^{\mathrm{S}}$ and $\mathbf{D}^{\mathrm{dCSI}}$]

Using Theorem \ref{thm: inner-bound 2user IC d-CSI and op fb} above and \cite[Theorem 2]{Vaze_Varanasi_delay_MIMO_IC}, we observe that $\mathbf{D}^{\mathrm{S}} \not= \mathbf{D}^{\mathrm{dCSI}}$ if and only if Conditions $1$ or $2$ hold. In other words, in the presence of delayed CSIT, output feedback helps in improving the DoF region only when Conditions 1 or 2 hold.
\end{remark}

%\begin{remark}
%The last theorem proves that for a large class of MIMO ICs, it is sufficient to have just CSI feedback, instead of having CSI and output feedback, both. This conclusion is useful from a practical perspective because output feedback is more expensive than CSI feedback since the former needs to be done on the per-time-slot basis while the latter needs to be performed once per coherence block. Thus, our results proves the futility of having output feedback in presence of CSI feedback for a large class of MIMO ICs, and the same time, asserts that the same conclusion is just not true for the MIMO ICs not belonging to this class.
%\end{remark}

\begin{remark}
Using Lemma \ref{lem: DoF region 2user IC p-CSI and op-fb}, %which is an extension of \cite[Corollary 11]{Chiachi-Jafar},
we observe that output feedback can not enhance the DoF region when there is instantaneous CSIT. In contrast, output feedback improves the DoF region when there is just delayed CSIT.
\end{remark}

%\begin{remark}
%As pointed out in the last section, our result is the first instance where feedback is shown to improve the DoF with delayed CSIT.
%
%
%Using the result of \cite{Vaze-Varanasi-delay-MIMOBC}, we observe that the DoF region of the $2$-user MIMO broadcast channel with Shannon feedback is equal to its DoF region with just delayed CSIT. In other words, output feedback, when present in addition to delayed CSIT, fails to improve the DoF region of the MIMO broadcast channel. However, in light of results obtained here and in \cite{Vaze_Varanasi_delay_MIMO_IC}, we now know that the same is not the case with the $2$-user MIMO IC.
%\end{remark}

The next two corollaries extend the above results to MIMO ICs with limited Shannon feedback of Type I and Type II, just output feedback and with designable Shannon feedback.

\vskip 1em
\begin{corollary} \label{cor: limited and designable Shannon feedback journal}
For the MIMO IC with i.i.d. Rayleigh fading, we have
\[
\mathbf{D}^{\rm \mathit{l}S1} = \mathbf{D}^{\rm \mathit{l}S2} = \mathbf{D}^{\rm \mathit{d}S} = \mathbf{D}^{\rm S}.
\]
\end{corollary}
\begin{IEEEproof}
See Appendix \ref{app: proof of cor: limited and designable Shannon feedback journal}.
\end{IEEEproof}
\vskip 1em
\begin{corollary} \label{cor: just output feedback Shannon journal}
For the MIMO IC with i.i.d. Rayleigh fading, we have
\[
\mathbf{D}^{\rm op} = \mathbf{D}^{\rm S},
\]
if neither of the following two inequalities hold: $\min(M_1,N_1) > N_2 > M_2$ and $\min(M_2,N_2) > M_1 > N_1$.
\end{corollary}
\begin{IEEEproof}
See Appendix \ref{app: proof of cor: just output feedback Shannon journal}
\end{IEEEproof}
\vskip 1em
Thus, the above corollary yields the DoF region with output feedback for a large class of MIMO ICs. When one of the above two conditions holds the DoF region with output feedback is not known.

Following the submission of a conference version of this work, and simultaneously with its publication in \cite{Vaze-MV-Shannonfb:Allerton11}, Tandon et. al. reported  the DoF region for limited Shannon feedback of Type II in \cite{tandon_Shannon_feedback_2011_arxiv}.
%While being in the final stages of writing this paper, the authors became aware of a recent work \cite{tandon_Shannon_feedback_2011_arxiv}. This work considers just the case of limited Shannon feedback of Type 2. In contrast, our results are applicable to a wide variety of settings, as pointed out earlier.

\subsection{Retrospective Interference Alignment with Shannon Feedback} \label{subsec: IA example dCSI_opfb 2user IC}

\begin{figure}[b]
\begin{centering}
\includegraphics[bb=0bp 55bp 720bp 530bp,scale=0.425]{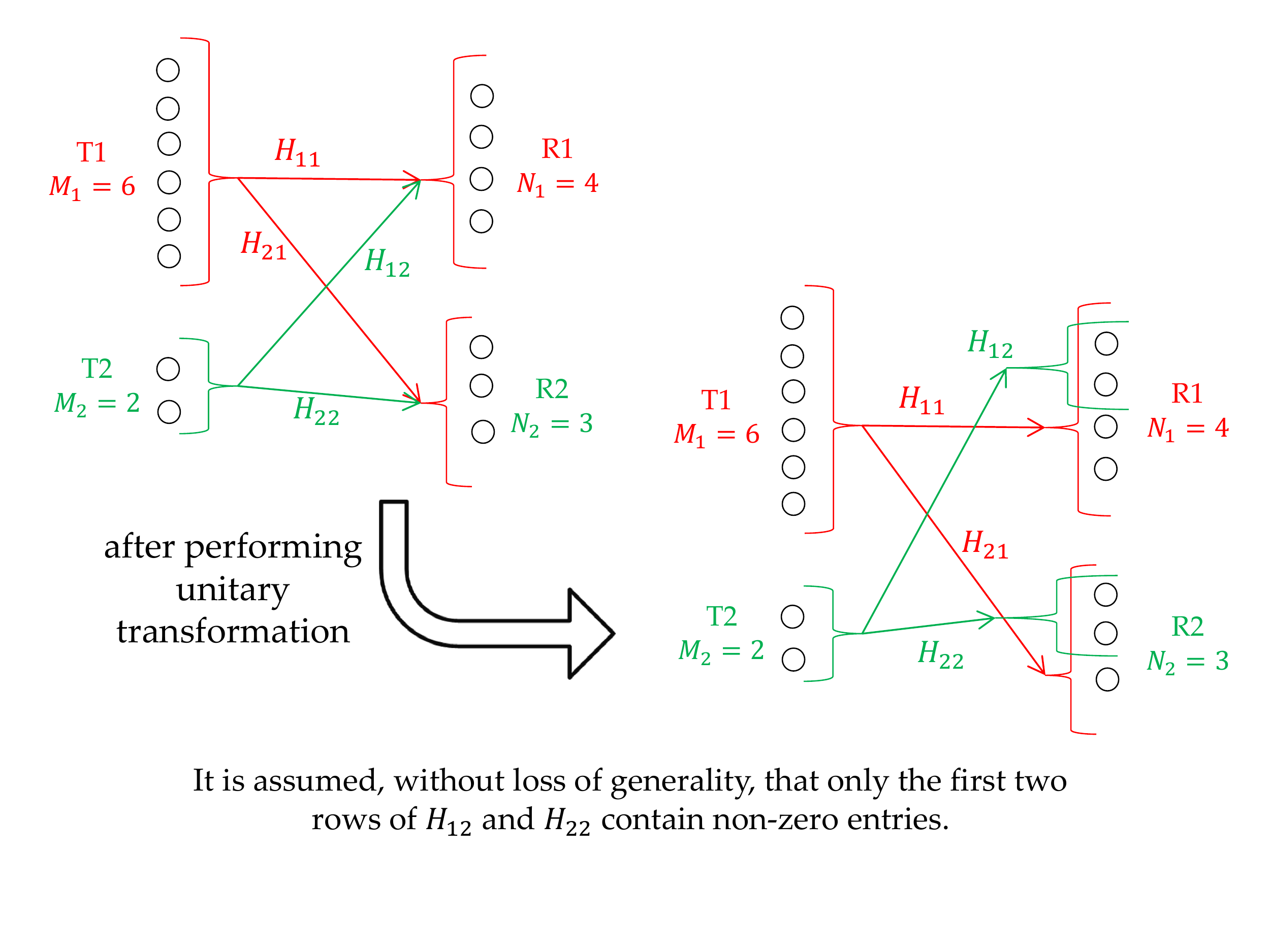}
\par\end{centering}
\caption{The $(6,2,4,3)$ MIMO IC Considered in Section \ref{subsec: IA example dCSI_opfb 2user IC}} \label{fig: example IC dCSI_opfb 2user IC}
\end{figure}

With Theorem \ref{thm: inner-bound 2user IC d-CSI and op fb} in hand, we know that $\mathbf{D}^{\mathrm{S}} \supset \mathbf{D}^{\mathrm{dCSI}}$ in general. However, the proof of this theorem is involved, and therefore, we provide an example in which $\mathbf{D}^{\mathrm{S}} \neq \mathbf{D}^{\mathrm{dCSI}}$ by demonstrating that a point outside $ \mathbf{D}^{\mathrm{dCSI}} $ can be achieved with Shannon feedback. The proof that this scheme works is based on a series of simple propositions.%, the outlines of the proofs of which are relegated to Appendix \ref{app: proofs of propositions in IA example Shannon journal}. %However, the main idea behind this achievability scheme can be understood even without these proofs.

In particular, we consider the $(6,2,4,3)$ MIMO IC shown in Fig. \ref{fig: example IC dCSI_opfb 2user IC}. For this IC, the DoF regions with just delayed CSIT and with Shannon feedback are plotted in Fig. \ref{fig: DoF region comparison dCSI+opfb 2user IC}, from which we observe that the former is strictly smaller than the latter. Moreover, it is clear from Fig. \ref{fig: DoF region comparison dCSI+opfb 2user IC} that when $d_2 = 2$, $d_1 \leq \frac{5}{3}$ with delayed CSIT. Here, we prove the achievability the DoF pair $(\frac{12}{7}, 2)$ with Shannon feedback, which establishes that $\mathbf{D}^{\mathrm{S}} \neq \mathbf{D}^{\mathrm{dCSI}}$ since $\frac{12}{7} > \frac{5}{3}$.

Toward this end, we show that by coding over $7$ times slots, $12$ and $14$ DoF can be achieved for the first and the second transmit-receive pairs, respectively. In our scheme, T2 transmits $2$ data symbols (DSs) intended for R2 over each time slot and thus a total of $14$ DSs are sent; whereas T1 transmits $6$ DSs intended for R1 at $t = 1$ and $t=4$. Further, at $t = 7$, we show that desired DSs can be successfully decoded by both receivers.

Consider first a transformation which simplifies the description of our scheme. At time $t$, the $i^{th}$ receiver can compute a unitary matrix $U_{i2}(t)$ such that it is deterministic function of $H_{i2}(t)$ and the bottom $(N_i - 2)$ rows of the transformed matrix $U_{i2}(t) H_{i2}(t) $ consist only of zeros. Using it, the $i^{th}$ receiver evaluates the transformed output $U_{i2}(t) Y_i(t) $. Henceforth, the transformed quantities $ U_{i2}(t) H_{i2}(t) $ and $ U_{i2}(t) Y_i(t) $ are denoted simply as $ H_{i2}(t) $ and $ Y_i(t) $, respectively. Evidently, the transmit signal $X_2(t)$ affects only the first two entries of (the transformed) ${Y}_i(t)$. Hence, throughout this subsection, we assume without loss of generality that the bottom $(N_i - 2)$ rows of $H_{i2}(t)$ consist only of zeros for all $t$, and thus, the signal $X_2(t)$ can affect only the first two antennas of R1 and R2 (see also Fig. \ref{fig: example IC dCSI_opfb 2user IC}).

\begin{figure}[t]
\begin{centering}
\includegraphics[scale=0.5]{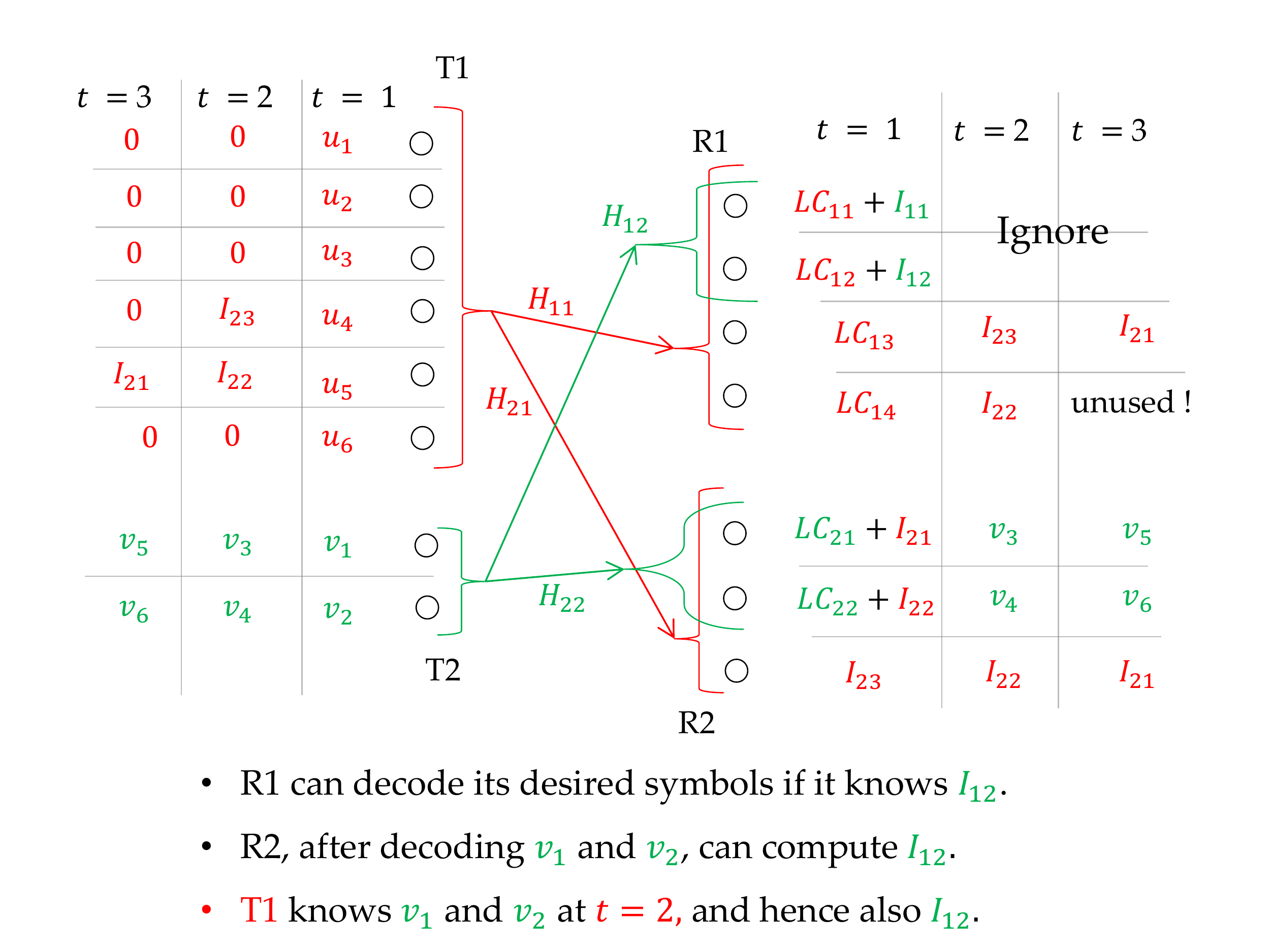}
\par\end{centering}
\caption{IA scheme for Achieving $(\frac{12}{7},2)$ with Shannon Feedback over the $(6,2,4,3)$ MIMO IC: $t=1$ to $t = 3$} \label{fig: IA example Shannon journal t=1-3}
\end{figure}

Consider the operation of the scheme at $t = 1$; see also Fig. \ref{fig: IA example Shannon journal t=1-3}. At this time, T1 transmits i.i.d. complex Gaussian DSs $\big\{u_i\}_{i=1}^6$ intended for R1, while T2 sends i.i.d. complex Gaussian DSs $v_1$ and $v_2$ for R2. Thus, the transmit signals are formed (for a vector $V_i$, $V_{ij}$ denotes its $j^{th}$ entry) as follows: $X_{1i}(1) = u_i$ $\forall$ $i \in [1:6]$ and $X_{2j}(1) = v_j$, $j=1,2$. The received signals at R1 and R2 can be written as follows (note that since the additive noises do not alter a DoF result, they are ignored) with desired and interfering linear combinations of data symbols defined using the symbols $LC$ and $I$, respectively, so that
\begin{eqnarray*}
Y_{1i}(1) & = & H_{1i1}(1) X_1(1) + H_{1i2}(1) X_2(1),  \qquad \qquad i \in [1:4] \\
          & = & \underbrace{ H_{1i1}(1) \begin{bmatrix} u_1^* & u_2^* & \cdots & u_6^* \end{bmatrix}^* }_{\define ~ {\rm LC}_{1i}}  + \underbrace{ H_{1i2}(1) \begin{bmatrix} v_1^* & v_2^* \end{bmatrix}^*}_{\define ~ {\rm I}_{1i}} \\
Y_{2i}(1) & = & H_{2j2}(1) X_2(1) + H_{2j1}(1) X_1(1) \qquad \qquad j \in [1:2] \\
          & = & \underbrace{ H_{2j2}(1) \begin{bmatrix} v_1^* & v_2^* \end{bmatrix}^* }_{\define ~ {\rm LC}_{2j}}  + \underbrace{ H_{2j1}(1) \begin{bmatrix} u_1^* & u_2^* & \cdots & u_6^* \end{bmatrix}^*}_{\define ~ {\rm I}_{2j}},
\end{eqnarray*}
and moreover, since we have assumed without loss of generality that the bottom $(N_i - 2)$ rows of $H_{i2}(t)$ consist only of zeros $\forall$ $t$ and $i$, we have ${\rm I}_{13} = {\rm I}_{14}  = {\rm LC}_{23} = 0$ (see also Fig. \ref{fig: IA example Shannon journal t=1-3}). Thus,
\begin{eqnarray*}
\begin{matrix}
Y_{1i}(1) & = & {\rm LC}_{1i} + {\rm I}_{1i} & \qquad & i \in [1:2], & \\
Y_{1i}(1) & = & {\rm LC}_{1i} & \qquad & i \in [3:4],  &  \\
Y_{2j}(1) & = & {\rm LC}_{2j} + {\rm I}_{2j} & \qquad & j \in [1:2], & \mbox{ and }\\
Y_{23}(1) & = & {\rm I}_{23}. & & &
\end{matrix}
\end{eqnarray*}

At time $t = 1$, both receivers encounter interference, and therefore, can not decode their desired data symbols. Moreover, the interference at a given antenna of a receiver is the linear combination of the DSs sent by its unpaired transmitter. Thus, each transmitter can compute the past interference encountered by its unpaired receiver using just delayed CSI, as stated in the following proposition.
\begin{proposition} \label{prop: compute interf. at unpaired Shannon journal example}
T1, at time $t = 2$, can compute ${\rm I}_{2j}$ $\forall$ $j \in [1:3]$ using just delayed CSI.
\end{proposition}
\begin{IEEEproof}
T1 knows DSs $u_i$'s; and time $t = 2$, and it also knows $H_{2j1}(1)$ because of delayed CSIT knowledge and hence ${\rm I}_{2j} =  H_{2j1}(1) \begin{bmatrix} u_1^* & u_2^* & \cdots & u_6^* \end{bmatrix}^*$.
%See Appendix \ref{app: proofs of propositions in IA example Shannon journal prop: compute interf. at unpaired Shannon journal example}.
\end{IEEEproof}

Hence, at $t = 2$ and $t = 3$, T1 transmits ${\rm I}_{2j}$ $\forall$ $j \in [1:3]$ as shown in Fig. \ref{fig: IA example Shannon journal t=1-3}, while T2 continues to transmit $2$ new DSs intended for R2. In particular, the transmit signals are given as
\begin{eqnarray*}
X_{1i}(2) & = & X_{1i}(3) = 0 ~ ~ \forall ~ ~ i \in [1:3], \\
X_{14}(2) & = & I_{23}, \qquad X_{15}(2) = I_{22}, \qquad X_{16}(2) = 0 \\
X_{14}(3) & = & I_{23}, \qquad X_{15}(3) = X_{16}(3) = 0 \\
X_{21}(2) & = & v_3, \qquad X_{22}(2) = v_4 \\
X_{21}(3) & = & v_5, \qquad X_{22}(3) = v_6,
\end{eqnarray*}
where $v_3$, $\cdots$, $v_6$ are i.i.d. complex Gaussian DSs intended for R2. Consider now the decoding operation at the receivers, starting with R2. The following proposition states that R2 can decode the desired DSs $v_1$, $\cdots$, $v_6$ at time $t = 3$.
\begin{proposition} \label{prop: example Shannon R2 decoding successful block 1}
At time $t = 2$, R2 can decode $v_3$, $v_4$, and $I_{22}$. At time $t = 3$, R2 can decode $v_5$, $v_6$, and $I_{21}$. After determining $I_{22}$ and $I_{21}$, R2 can evaluate $v_1$ and $v_2$, and thus, at $t = 3$, decoding is successful at R2.
\end{proposition}
\begin{IEEEproof}
At time $t = 1$, R2 knows $I_{23}$. Hence, it can subtract the contribution due to $I_{23}$ from $Y_2(2)$. Thus, equivalently, for R2, only $3$ transmit antennas sending a non-zero signal at this time. Therefore, R2, via simple channel inversion, can determine $v_3$, $v_4$, and $I_{22}$.
That at time $t = 3$, R2 can decode $v_5$, $v_6$, and $I_{21}$ follows similarly. After knowing the values of $I_{21}$ and $I_{22}$, R2 can evaluate $Y_{21}(1) - I_{21} = {\rm LC}_{21}$, and similarly, ${\rm LC}_{22}$. In other words, at $t = 3$, it can obtain $2$ linear combinations (LCs) of $v_1$ and $v_2$. Thus, it can decode $v_1$ and $v_2$.
\end{IEEEproof}

Consider now the case of R1. At $t = 3$, as per the next proposition, R1 knows $I_{21}$, $I_{22}$, and $I_{23}$. Since these are linear combinations of $u_1$, $\cdots$, $u_6$, they are useful for R1.
\begin{proposition} \label{prop: R1 gets interf at R2 ex Shannon journal}
R1 can determine the values of $I_{22}$ and $I_{23}$ at time $t = 2$, and that of $I_{21}$ at time $t = 3$.
\end{proposition}
\begin{IEEEproof}
R1 can simply ignore the first two receive antennas, which experience interference due to the signal of T2. Then, using the last two antennas, it can compute the required symbols using channel inversion.
\end{IEEEproof}

Thus, at $t = 3$, R1 gets $5$ linear combinations, namely, ${\rm LC}_{23}$, ${\rm LC}_{24}$, $I_{21}$, $I_{22}$, and $I_{23}$, of $6$ desired DSs. Thus, it needs one more useful linear combination for successful decoding. Consider the next proposition.
\begin{proposition} \label{prop: R1 knows interf at itself ex Shannon journal}
Given that R1 knows the values of ${\rm LC}_{23}$, ${\rm LC}_{24}$, $I_{21}$, $I_{22}$, and $I_{23}$, it can decode the six symbols $u_1$, $u_2$, $\cdots$, $u_6$, provided it knows $I_{12}$.
\end{proposition}
\begin{IEEEproof}
If R1 knows ${\rm LC}_{23}$, ${\rm LC}_{24}$, $I_{21}$, $I_{22}$, and $I_{23}$, then it is sufficient for it to know one more linear combination of $u_1$, $u_2$, $\cdots$, $u_6$, which it can compute using $I_{12}$ as follows: $Y_{12}(1) - I_{12} = {\rm LC}_{12}$ is a linear combination of R1's desired symbols.
\end{IEEEproof}

Hence, it is sufficient to communicate the value of $I_{12}$ to R1. Consider the following proposition.
\begin{proposition} \label{prop: T1 knows interf R1 ex Shannon journal}
T1 can compute $I_{12}$ at $t = 2$ using Shannon feedback, but not using just delayed CSI. Moreover, R2 knows $I_{12}$ at time $t = 3$.
\end{proposition}
\begin{IEEEproof}
Because of Shannon feedback, T1 knows $Y_2(1)$, $H_{21}(1)$, and $H_{22}(1)$ at time $t = 2$ by virtue of Shannon feedback. Since it knows $X_1(1)$ by default, it can compute
\[
\Big( H_{22}(1) \Big)^{\dagger} \Big\{ Y_2(1) - H_{21}(1) X_1(1)\Big\} = \Big( H_{22}(1) \Big)^{\dagger} H_{22}(1) X_2(1) = X_2(1)
\]
at time $t=2$, where $\Big( H_{22}(1) \Big)^{\dagger}$ is the pseudo-inverse of $H_{22}(1)$ \cite{Horn-Johnson}. Subsequently, it can evaluate $I_{12}$, which is a linear combination of $v_1$ and $v_2$.
R2, after decoding $v_1$ and $v_2$ at time $t = 3$, can compute $I_{12}$ since it knows all channel matrices.
\end{IEEEproof}
Thus, in light of this proposition, T1 can convey $I_{12}$ to R1 at time $t = 6$ without interfering with decoding at R2.

\begin{figure}[t]
\begin{centering}
\includegraphics[scale=0.5]{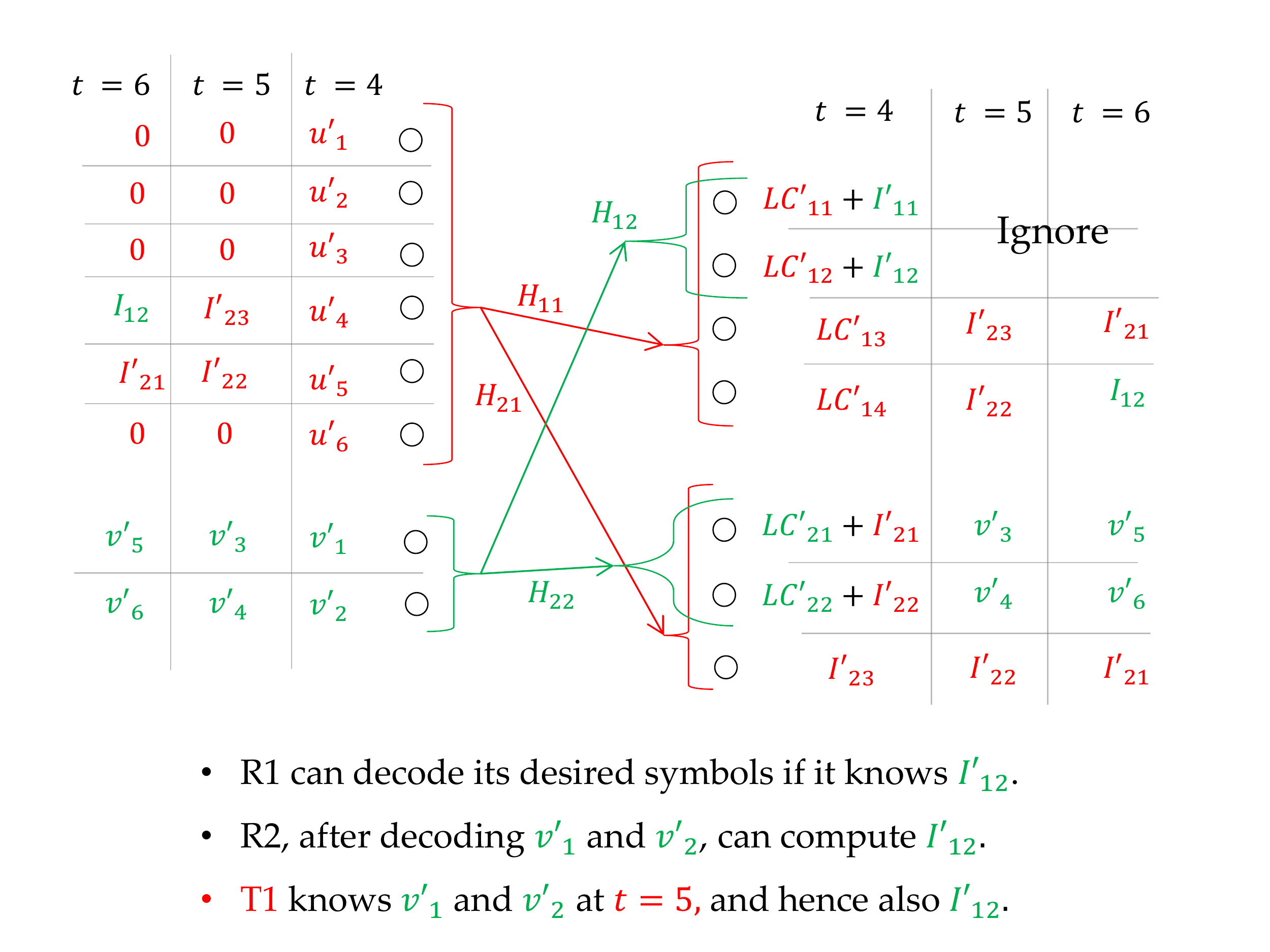}
\par\end{centering}
\caption{IA scheme for Achieving $(\frac{12}{7},2)$ with Shannon Feedback over the $(6,2,4,3)$ MIMO IC: $t=4$ to $t = 6$} \label{fig: IA example Shannon journal t=4-6}
\end{figure}

Consider the next block of $3$ time slots, i.e., for $t =4,5,6$. The scheme here is identical to that for the first three time slots; see Fig. \ref{fig: IA example Shannon journal t=4-6}.
At time $t = 4$, T1 transmits DSs $u'_1$, $u'_2$, $\cdots$, $u'_6$ intended for R1. T2, on the other hand, transmits DSs $v'_1$, $v'_2$, $\cdots$, $v'_6$ for R2. Here, the superscript prime is used to indicate the quantities that are specific to this block of three time slots. The only change in the transmission scheme is that at time $t = 6$, T1 transmits $I_{12}$. Consider the following propositions which will be used to describe decoding at the receivers.

\begin{proposition} \label{prop: example Shannon R2 decoding successful block 2}
Consider receiver R2: (a) at time $t = 5$, R2 can decode $v'_3$, $v'_4$, and $I'_{22}$ (b) at time $t = 6$, R2 can decode $v'_5$, $v'_6$, and $I'_{21}$ and (c) after determining $I'_{22}$ and $I'_{23}$, R2 can evaluate $v'_1$ and $v'_2$, and thus, at $t = 6$, R2 can perform successful decoding.
\end{proposition}
\begin{IEEEproof}
The proofs of Parts (a) and (c) similar to those of Parts (i) and (iii) of Proposition \ref{prop: example Shannon R2 decoding successful block 1}. Further, Part (b) follows from the proof of Proposition \ref{prop: example Shannon R2 decoding successful block 1}(ii) by noting that $I_{12}$ is known to R1 at time $t = 3$.
\end{IEEEproof}

\begin{figure}[t]
\begin{centering}
\includegraphics[bb=0bp 60bp 720bp 450bp,clip,scale=0.5]{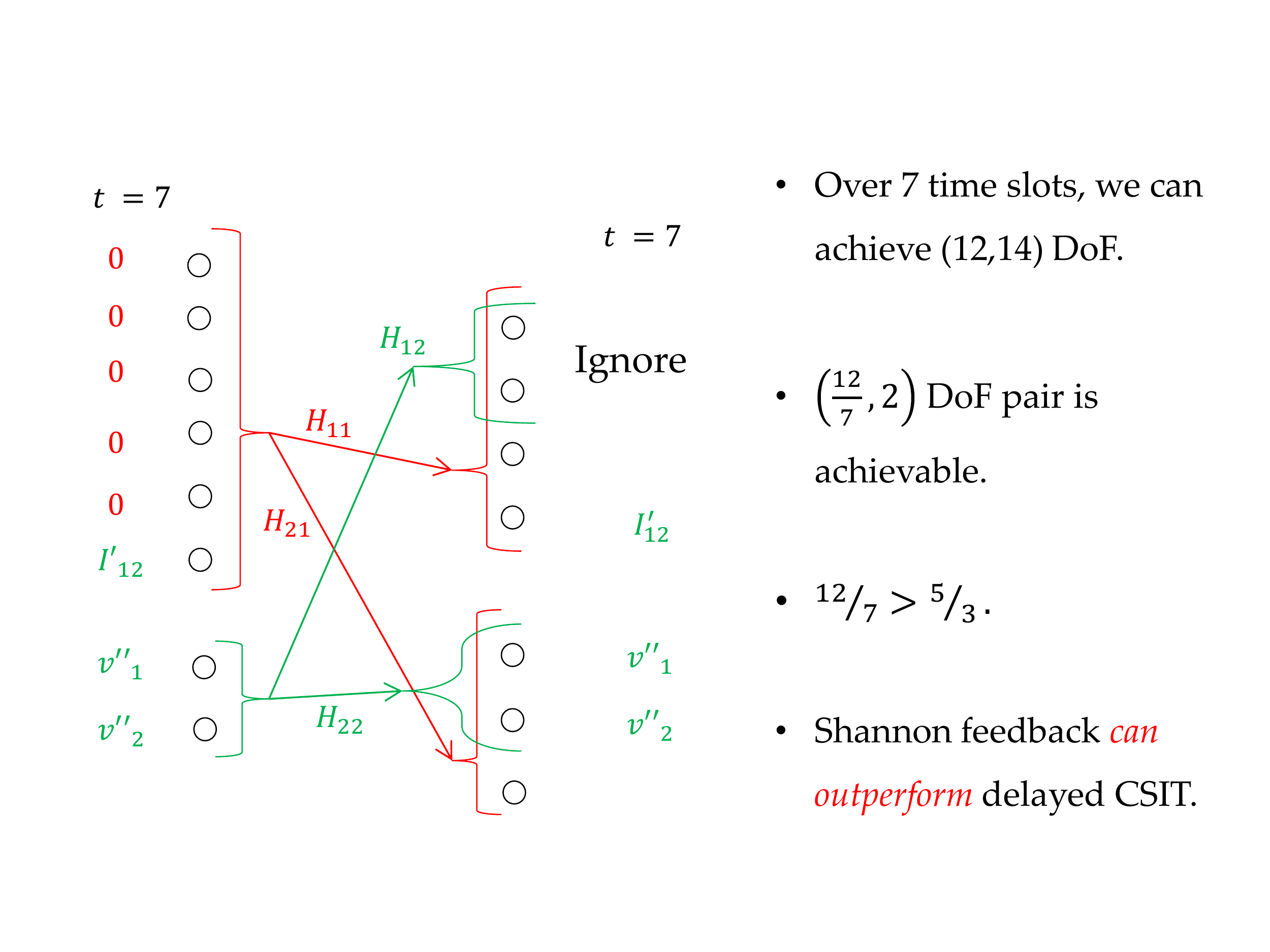}
\par\end{centering}
\caption{IA scheme for Achieving $(\frac{12}{7},2)$ with Shannon Feedback over the $(6,2,4,3)$ MIMO IC: $t=7$} \label{fig: IA example Shannon journal t=7}
\end{figure}

\begin{proposition} \label{prop: R1 decoding block 2 ex Shannon journal}
Consider receiver R1: (a) R1 knows $I_{12}$ at time $t = 6$, from which it can decode $u_1$, $u_2$, $\cdots$, $u_6$, (b) R1 can compute $I'_{22}$ and $I'_{23}$ at time $t = 5$, and $I'_{21}$ at time $t = 6$ (c) if R1 is conveyed the value of $I'_{12}$ at time $t = 7$, it can decode $u'_1$, $u'_2$, $\cdots$, $u'_6$ and (d) T1 knows $I'_{12}$ at time $t = 5$.
\end{proposition}
\begin{IEEEproof}
The four parts follow respectively from Propositions \ref{prop: R1 knows interf at itself ex Shannon journal}, \ref{prop: R1 gets interf at R2 ex Shannon journal}, \ref{prop: R1 knows interf at itself ex Shannon journal}, and \ref{prop: T1 knows interf R1 ex Shannon journal}.
\end{IEEEproof}

Thus, as per the two propositions, at $t = 6$, R2 can decode all symbols sent to it until that time, whereas R1 can do the same if it is delivered the value of $I'_{12}$ at time $t = 7$. Next, consider $t = 7$. As shown in Fig. \ref{fig: IA example Shannon journal t=7}, T1 transmits just $I'_{12}$, while T2 sends two new data symbols $v^"_1$ and $v^"_2$. It is easy to show that R1 can decode $I'_{12}$, whereas R2 can decode $v^"_1$ and $v^"_2$. Hence, as desired, we can achieve a DoF pair $(12,14)$ over $7$ symbol times.

\begin{figure}[t]
\begin{centering}
\includegraphics[bb=0bp 40bp 720bp 480bp,clip,scale=0.5]{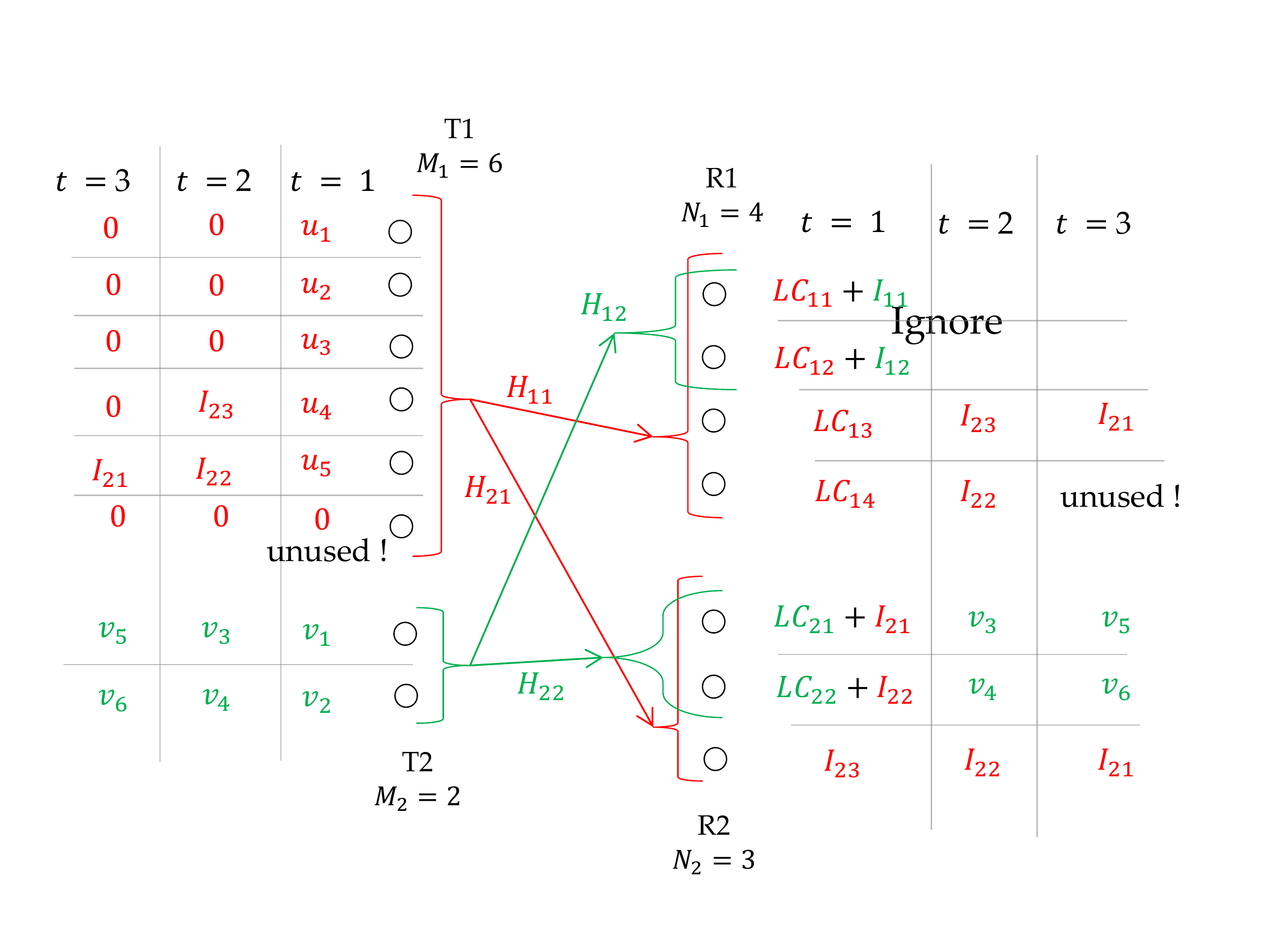}
\par\end{centering}
\caption{IA scheme for Achieving $(\frac{5}{3},2)$ with Delayed CSIT over the $(6,2,4,3)$ MIMO IC} \label{fig: delayed CSIT IA example Shannon journal t=1-3}
\end{figure}

It is instructive to compare the above Shannon feedback scheme with the delayed-CSIT coding scheme of \cite{Vaze_Varanasi_delay_MIMO_IC} that can only achieve the pair $(\frac{5}{3},2)$ over the $(6,2,4,3)$ MIMO IC as illustrated in Fig. \ref{fig: delayed CSIT IA example Shannon journal t=1-3} in terms of the notation introduced earlier in this sub-section. In this latter case, by coding over $3$ time slots, we achieve $5$ and $6$ DoF for the two transmit-receive pairs, respectively. Note that T1 can not determine $I_{12}$ (and $I'_{12}$) with just delayed CSIT, and thus only $5$ linear combinations can be delivered to R1 over a span of $3$ time slots. Hence, T1 transmits only $5$ DSs for R1 at time $t = 1$, which R1 can decode at $t = 3$. Except for this difference (compare Figs. \ref{fig: IA example Shannon journal t=1-3} and \ref{fig: delayed CSIT IA example Shannon journal t=1-3}), this coding scheme is identical to the Shannon feedback coding scheme.

Note that under this delayed-CSIT scheme, the sixth antenna of R1 is never used. Moreover, one of the last two antennas of R1, say, the fourth, remains unused, although it never experiences interference. In other words, in the delayed-CSIT scheme, some of the resources are not utilized. Shannon feedback on the other hand permits the exploitation of these resources -- the sixth antenna of T1 is used with Shannon feedback and both interference-free antennas of R1 are used under the Shannon-feedback scheme at $t = 4,5,6$ -- thereby outperforming delayed CSIT feedback. % Thus, one may conclude that Shannon feedback results in a better utilization of all available resources, which manifests itself in terms of the DoF improvement.

\subsection{Comparison of IA with Shannon Feedback and IA with Delayed CSIT} \label{subsec: intuition behind results dCSI_opfb 2user IC}

In the Shannon-feedback coding scheme of the previous sub-section (and more generally of Section \ref{sec: proof of thm: inner-bound 2user IC d-CSI and op fb}), one may observe that the following two types of techniques are used to achieve IA:
\begin{enumerate}
\item Since the interference at a given receiver is a linear combinations of the DSs sent by its unpaired transmitter, each transmitter, using delayed CSIT, can evaluate and then transmit the interference seen in the past by its {\em unpaired} receiver to convey new useful linear combinations to its paired receiver without creating any new additional interference to the unpaired receiver.
\item Equipped with the knowledge of past channel outputs and past channel matrices, each transmitter can compute and transmit the interference encountered in the past by its {\em paired} receiver to provide an opportunity to its paired receiver to cancel the past interference while conveying useful information to its unpaired receiver.
\end{enumerate}

Note that the DoF-region-optimal IA-based achievability schemes developed in \cite{Vaze_Varanasi_delay_MIMO_IC} for the MIMO IC with {\em just} delayed CSIT make use of the first technique but not the second one, because the latter is feasible only in the presence of output feedback. Output feedback with delayed CSIT on the other hand enables each transmitter to compute the past transmit signal of the other transmitter which introduces partial transmitter cooperation which is infeasible when there is just delayed CSIT; remarkably, this transmitter cooperation reveals all available signaling dimensions and achieves the DoF gains promised by Shannon feedback.

Note that while transmitter cooperation is induced by output feedback regardless of whether there is instantaneous CSIT or delayed CSIT, it is  only in the case of delayed CSIT that such cooperation provides a DoF-region improvement. With instantaneous CSIT, the transmit signals can be suitably beamformed to cause minimal interference at the receivers. With delayed CSIT however, transmit beamforming can not be employed and hence the receivers experience a relatively high amount of interference so that output feedback is more effective.

%To understand this, note that with instantaneous CSIT, the transmit signals can be appropriately beamformed to cause minimal interference to the receivers, which is infeasible with just delayed CSIT, and therefore, the cooperation introduced by output feedback is more valuable when there is just delayed CSIT than it is when there is instantaneous CSIT.

%the interference encountered presently by the $i^{th}$ receiver, which is equal to the linear combinations of the DSs intended for the $j^{th}$ receiver, $j \not= i$, can be perfectly evaluated and therefore transmitted by the $j^{th}$ transmitter at a future time instant to enable the $i^{th}$ receiver to cancel a part of the pre-encountered interference, while simultaneously delivering a new useful linear combination to the $j^{th}$ receiver without producing any additional interference at the $i^{th}$ receiver.
%
%
% transmitter can compute the current transmit signal of the other transmitter at a later time instant. Thus, a given transmitter can retransmit the past transmit signal of the other transmitter, which on the one hand, provides its paired receiver an opportunity to eliminate a part of the interference that it has experienced earlier, while, on the other, delivers useful linear combinations to its unpaired receiver.

\section{Proof of Theorem \ref{thm: outer-bound 2user IC d-CSI and op fb}} \label{sec: proof of thm: outer-bound 2user IC d-CSI and op fb}

If $(d_1,d_2) \in \mathbf{D}^{\mathrm{S}}$, then $(d_1,d_2) \in \mathbf{D}^{\mathrm{iCSI \& op}}$. Therefore, by Lemma \ref{lem: DoF region 2user IC p-CSI and op-fb}, bounds $L_{01}$, $L_{02}$, and $L_3$ must hold at any $(d_1,d_2) \in \mathbf{D}^{\mathrm{S}}$. Now, note that $L_1$ and $L_2$ are symmetric counterparts of each other (i.e., any one of them can be obtained from the other by changing the user ordering). Hence, it is sufficient to prove that $L_1$ holds, which is the goal of the remainder of this section.

\begin{figure}[t] \centering
\includegraphics[bb=0bp 185bp 540bp 650bp,clip, height=3in, width=3in]{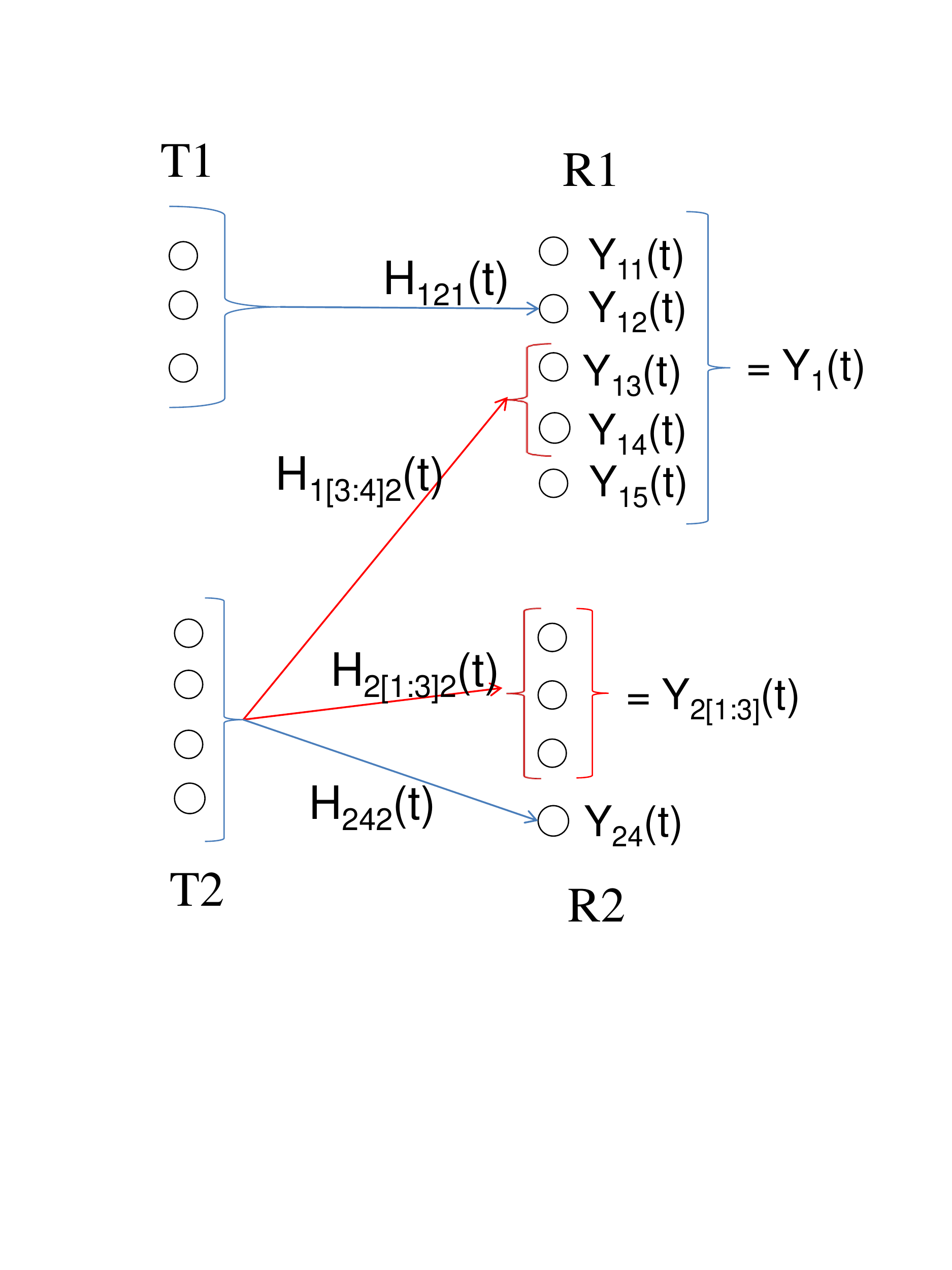}
\caption{Illustration of notation.} \label{fig: notation_used Shannon journal}
\end{figure}

Before we prove that $L_1$ is an outer-bound, we introduce some further notation used henceforth in the paper.

\emph{Notation: } The set of four channel matrices at time $t$ is denoted by $H(t)$, i.e., $H(t) = \big\{ H_{ij}(t) \big\}$ where $i,j \in \{1,2\}$. For integers $n_1$ and $n_2$, if $n_1 \leq n_2$, $[n_1:n_2] =  \{n_1,n_1+1, \cdots, n_2\}$; whereas if $n_1 > n_2$, then $[n_1: n_2]$ denotes the empty set. For a random variable $X(t)$, $X([n_1:n_2]) = \{X(t)\}_{t=n_1}^{n_2}$ if $n_1 \leq n_2$, whereas $X([n_1:n_2])$ denotes an empty set if $n_1 > n_2$. Further, for $n \geq 1$, $\overline{X}(n) = X([1:n])$. For the received signal $Y_i(t)$ and the channel matrix $H_{ik}(t)$, the $j^{th}$ entry and the $j^{th}$ row are denoted respectively by $Y_{ij}(t)$ and $H_{ijk}(t)$. Further, whenever $n_1 \leq n_2$ and $n_3 \leq n_4$, $Y_{i[n_1:n_2]}(t) = \{Y_{ij}(t)\}_{j=n_1}^{n_2}$, $Y_{i[n_1:n_2]}([n_3:n_4]) = \big\{ \{Y_{ij}(t)\}_{j=n_1}^{n_2} \big\}_{t=n_3}^{n_4}$, $H_{i[n_1:n_2]j}(t)$ is the channel matrix from $j^{th}$ transmitter to channel outputs $Y_{i[n_1:n_2]}(t)$ (see Fig. \ref{fig: notation_used Shannon journal}); however, if $n_1 > n_2$ and/or $n_3>n_4$, then $Y_{i[n_1:n_2]}(t)$ and $Y_{i[n_1:n_2]}([n_3:n_4])$ denote empty sets. Moreover, for $n \geq 1$, $\overline{Y}_{i[n_1:n_2]}(n) = Y_{i[n_1:n_2]}([1:n])$. Finally, $o(\log_2 P)$ denotes any real-valued function $x(P)$ of $P$ such that $\lim_{P \to \infty} \frac{x(P)}{\log_2 P} = 0$.

We will show that the bound $L_1$ must hold at any $(d_1,d_2) \in \mathbf{D}^{\mathrm{S}}$.
We first apply Fano's inequality to upper-bound the rates achievable for the two users starting below with $R_2$.
\begin{eqnarray}
n R_2 & \leq & I \Big( \mathcal{M}_2 ; \overline{Y}_2(n) \Big| \overline{H}(n) \Big) + n \epsilon_n \\
& = & h \Big(\overline{Y}_2(n) \Big| \overline{H}(n) \Big) - h \Big(\overline{Y}_2(n) \Big| \mathcal{M}_2, \overline{H}(n) \Big) + n \epsilon_n \\
& = & h \Big(\overline{Y}_2(n) \Big| \overline{H}(n) \Big) - \sum_{t=1}^n h \Big( Y_2(t) \Big| \overline{Y}_2(t-1), \mathcal{M}_2, \overline{H}(n) \Big) + n \epsilon_n \\
& \leq & h \Big(\overline{Y}_2(n) \Big| \overline{H}(n) \Big) - \sum_{t=1}^n h \Big( Y_2(t) \Big| \overline{Y}_2(t-1), \mathcal{M}_2, \overline{X}_2(t), \overline{H}(n) \Big) + n \epsilon_n, \label{eq: semifinal bound on R2 delayed CSI_op fb} \\
& = & h \Big(\overline{Y}_2(n) \Big| \overline{H}(n) \Big) - \sum_{t=1}^n h \Big( Y_2(t) \Big| \overline{Y}_2(t-1), \mathcal{M}_2, \overline{X}_2(t), \overline{H}(t) \Big) + n \epsilon_n, \label{eq: final bound on R2 delayed CSI_op fb}
\end{eqnarray}
where $\epsilon_n \to 0$ as $n \to \infty$; the inequality (\ref{eq: semifinal bound on R2 delayed CSI_op fb}) holds since conditioning reduces entropy \cite{CT}; and the equality in (\ref{eq: final bound on R2 delayed CSI_op fb}) follows on noting that random variables $\big\{ Y_2(t), \overline{Y}_2(t-1), \mathcal{M}_2, \overline{X}_2(t) \big\}$ are independent of $H([t+1:n])$.

We next use Fano's inequality at R1 assuming that it knows the received signal $Y_2(t)$ instantaneously and also the message $\mathcal{M}_2$ to obtain the following:
\begin{eqnarray}
n R_1 & \leq & I \Big( \mathcal{M}_1 ; \overline{Y}_2(n), \overline{Y}_1(n), \mathcal{M}_2 \Big| \overline{H}(n) \Big) + n \epsilon_n \nonumber \\
& = & I \Big( \mathcal{M}_1 ; \overline{Y}_1(n), \overline{Y}_2(n) \Big| \mathcal{M}_2,  \overline{H}(n) \Big) + n \epsilon_n \label{eq: stepI bound on R1 dCSI_opfb 2user IC} \\
& = & \sum_{t=1}^n h \Big( Y_1(t), Y_2(t) \Big| \overline{Y}_1(t-1), \overline{Y}_2(t-1), \mathcal{M}_2,  \overline{H}(n) \Big) \nonumber \\
&   & - \sum_{t=1}^n h \Big( Y_1(t), Y_2(t) \Big| \overline{Y}_1(t-1), \overline{Y}_2(t-1), \mathcal{M}_2,  \mathcal{M}_1, \overline{H}(n) \Big) + n \epsilon_n \label{eq: stepII bound on R1 dCSI_opfb 2user IC} \\
& = & \sum_{t=1}^n h \Big( Y_1(t), Y_2(t) \Big| \overline{Y}_1(t-1), \overline{Y}_2(t-1), \mathcal{M}_2, \overline{X}_2(t), \overline{H}(n) \Big) \nonumber \\
&   & - \sum_{t=1}^n h \Big( Y_1(t), Y_2(t) \Big| \overline{Y}_1(t-1), \overline{Y}_2(t-1), \mathcal{M}_2,  \mathcal{M}_1, \overline{X}_1(t), \overline{X}_2(t), \overline{H}(n) \Big) + n \epsilon_n  \label{eq: stepIII bound on R1 dCSI_opfb 2user IC} \\
& = & \sum_{t=1}^n h \Big( Y_1(t), Y_2(t) \Big| \overline{Y}_1(t-1), \overline{Y}_2(t-1), \mathcal{M}_2, \overline{X}_2(t), \overline{H}(t) \Big) \nonumber \\
&   & - \sum_{t=1}^n h \Big( Y_1(t), Y_2(t) \Big| \overline{Y}_1(t-1), \overline{Y}_2(t-1), \mathcal{M}_2,  \mathcal{M}_1, \overline{X}_1(t), \overline{X}_2(t), \overline{H}(t) \Big) + n \epsilon_n \label{eq: stepIIIb bound on R1 dCSI_opfb 2user IC} \\
& = & \sum_{t=1}^n h \Big( Y_1(t), Y_2(t) \Big| \overline{Y}_1(t-1), \overline{Y}_2(t-1), \mathcal{M}_2, \overline{X}_2(t), \overline{H}(t) \Big) \nonumber \\
&   & - \sum_{t=1}^n h \Big( W_1(t), W_2(t) \Big) + n \epsilon_n \label{eq: stepIV bound on R1 dCSI_opfb 2user IC}\\
& = & \sum_{t=1}^n \left\{ h \Big( Y_1(t), Y_2(t) \Big| \overline{Y}_1(t-1), \overline{Y}_2(t-1), \mathcal{M}_2, \overline{X}_2(t) \overline{H}(t) \Big) +  o(\log_2 P) + \epsilon_n\right\}, \label{eq: final bound on R1 delayed CSI_op fb}
\end{eqnarray}
where various steps follow because of the following reasons: the equality in \eqref{eq: stepI bound on R1 dCSI_opfb 2user IC} holds due to the independence of the two messages; equality (\ref{eq: stepII bound on R1 dCSI_opfb 2user IC}) holds because of the definition of the mutual information and the chain rule for the differential entropy; equality (\ref{eq: stepIII bound on R1 dCSI_opfb 2user IC}) follows by noting that the transmit signal $X_i(t)$ is a deterministic function of $\mathcal{M}_i$, $\overline{Y}_1(t-1)$, $\overline{Y}_2(t-1)$, and $\overline{H}(t)$; (\ref{eq: stepIIIb bound on R1 dCSI_opfb 2user IC}) holds since all the involved random variables are independent of $H([t+1:n])$; (\ref{eq: stepIV bound on R1 dCSI_opfb 2user IC}) holds because translation does not change differential entropy, and $W_1(t)$ and $W_2(t)$ are independent of $\overline{Y}_1(t-1)$, $\overline{Y}_2(t-1)$, $\mathcal{M}_2$,  $\mathcal{M}_1$, $\overline{X}_1(t)$, $\overline{X}_2(t)$, and $\overline{H}(t)$; the final equality holds since the noises are i.i.d. across time and their statistics are independent of $P$.

%Consider next the following lemma:
\begin{lemma} \label{lem: main inequality d-CSI+op-fb IC outer-bound}
Let $m_1 \define \min(M_1,N_1+N_2)$ and $m_2 \define \min(M_1,N_2)$. Then, for each $t \in [1:n]$, we have
\begin{eqnarray*}
\lefteqn{ \frac{1}{m_2} h \Big( Y_2(t) \Big| \overline{Y}_2(t-1), \mathcal{M}_2, \overline{X}_2(t), \overline{H}(t) \Big) } \\
&& {} \geq \frac{1}{m_1} h \Big( Y_1(t), Y_2(t) \Big| \overline{Y}_1(t-1), \overline{Y}_2(t-1), \mathcal{M}_2, \overline{X}_2(t), \overline{H}(t) \Big) + o(\log_2 P)
\end{eqnarray*}
where the term $o(\log_2 P)$ is constant with $n$.
\end{lemma}
\begin{IEEEproof}
See Section \ref{subsec: proof of lem: main inequality d-CSI+op-fb IC outer-bound}.
\end{IEEEproof}

Combining the bounds in (\ref{eq: final bound on R2 delayed CSI_op fb}), (\ref{eq: final bound on R1 delayed CSI_op fb}), and the one in Lemma \ref{lem: main inequality d-CSI+op-fb IC outer-bound}, we get
\begin{eqnarray*}
\frac{1}{m_2} R_2 & \leq & \frac{1}{m_2 \cdot n} h \Big(\overline{Y}_2(n) \Big| \overline{H}(n) \Big) + \frac{\epsilon_n}{m_2}  - \left\{ \frac{1}{m_1} R_1 - o(\log_2 P) - \epsilon_n \right\} \\
\Rightarrow \frac{R_2}{m_2} + \frac{R_1}{m_1} & \leq & \frac{\min(N_2,M_1+M_2)}{m_2}  \cdot \log_2 P + \epsilon_n \left( \frac{1}{m_2} + 1 \right) + o(\log_2 P),
\end{eqnarray*}
where the last inequality holds since the DoF of the point-to-point MIMO channel are equal to the minimum of the number of transmit and receive antennas. Since $\epsilon_n \to 0$ as $n \to \infty$, we now have
\begin{eqnarray*}
\frac{R_2}{m_2} + \frac{R_1}{m_1} & \leq & \frac{\min(N_2,M_1+M_2)}{m_2}  \cdot \log_2 P + o(\log_2 P) \\
\Rightarrow \frac{d_2}{m_2} + \frac{d_1}{m_1} & \leq & \limsup_{P \to \infty} \frac{R_2}{m_2} + \frac{R_1}{m_1} \leq \frac{\min(N_2,M_1+M_2)}{m_2}
\end{eqnarray*}
as desired.

\subsection{Proof of Lemma \ref{lem: main inequality d-CSI+op-fb IC outer-bound}} \label{subsec: proof of lem: main inequality d-CSI+op-fb IC outer-bound}

In the following two lemmas, it is shown that although the received signals $Y_1(t)$ and $Y_2(t)$ are $N_1$ and $N_2$ dimensional, respectively, only the first $m_1-m_2$ and $m_2$ entries of them are relevant as far as the current DoF analysis is concerned.
\begin{lemma} \label{lem: extracting essential part of term1 main inequality dCSI+opfb IC outer-bound}
If $m_2 = \min(M_1,N_2)$, we have the following:
\[
h \Big( Y_2(t) \Big| \overline{Y}_2(t-1), \mathcal{M}_2, \overline{X}_2(t), \overline{H}(t) \Big) \geq h \Big( Y_{2[1:m_2]}(t) \Big| \overline{Y}_2(t-1), \mathcal{M}_2, \overline{X}_2(t), \overline{H}(t) \Big) + o(\log_2 P),
\]
where the term $o(\log_2 P)$ is constant with $n$.
\end{lemma}
\begin{IEEEproof}
Follows from the techniques in \cite[Proof of Lemma 2]{Vaze_Varanasi_delay_MIMO_IC}.
\end{IEEEproof}

\begin{lemma} \label{lem: extracting essential part of term2 main inequality dCSI+opfb IC outer-bound}
If $m_1 = \min(M_1,N_1+N_2)$, then
\begin{eqnarray*}
\lefteqn{ h \Big( Y_1(t), Y_2(t) \Big| \overline{Y}_1(t-1), \overline{Y}_2(t-1), \mathcal{M}_2, \overline{X}_2(t), \overline{H}(t) \Big) } \\
&& {} \leq h \Big( Y_{1[1:m_1-m_2]}(t), Y_{2[1:m_2]}(t) \Big| \overline{Y}_1(t-1), \overline{Y}_2(t-1), \mathcal{M}_2, \overline{X}_2(t), \overline{H}(t) \Big) + o(\log_2 P).
\end{eqnarray*}
where the term $o(\log_2 P)$ is constant with $n$.
\end{lemma}
\begin{IEEEproof}
Follows from the techniques in \cite[Proof of Lemma 3]{Vaze_Varanasi_delay_MIMO_IC}.
\end{IEEEproof}

If $m_1 - m_2 = 0$, Lemma \ref{lem: main inequality d-CSI+op-fb IC outer-bound} holds trivially. Hence, in the following, we may consider without loss of generality that $m_1 > m_2$.

%\begin{remark}
%All differential entropy terms to be analyzed in this sub-section have $\overline{X}_2(t)$ within conditioning. Hence, throughout the remainder of this sub-section, it is assumed that $X_2(t_0) = 0_{M_2}$ $\forall$ $t_0 \in [1:t]$.
%\end{remark}

We now prove the following lemma which is critical in the proof of Lemma \ref{lem: main inequality d-CSI+op-fb IC outer-bound}.
\begin{lemma}\label{lem: statistical equivalence dCSI+opfb 2user IC outer-bound}
Let $Q(t) \define \big\{ \mathcal{M}_2, \overline{H}(t), \overline{Y}_2(t-1), \overline{X}_2(t) \big\}$. For an $i \in [1:m_2-1]$ and a $k \in [1:m_1-m_2]$, if $j = i+1$ and $l = k+1$, we have the following equalities:
\begin{eqnarray*}
h \Big( Y_{2i}(t) \Big| Q(t), Y_{2[1:i-1]}(t) \Big) & = & h \Big( Y_{2j}(t) \Big| Q(t), Y_{2[1:i-1]}(t) \Big); \\
h \Big( Y_{2m_2}(t) \Big| Q(t), Y_{2[1:m_2-1]}(t) \Big) & = & h \Big( Y_{11}(t) \Big| Q(t), Y_{2[1:m_2-1]}(t) \Big); \\
h \Big( Y_{1k}(t) \Big| Q(t), Y_{2[1:m_2]}(t), Y_{1[1:k-1]}(t) \Big) & = & h \Big( Y_{1l}(t) \Big| Q(t), Y_{2[1:m_2]}(t), Y_{1[1:k-1]}(t) \Big).
\end{eqnarray*}
\end{lemma}
\begin{IEEEproof}
It is sufficient to prove the first equality. Define $Y_2'(t) = Y_2(t) - H_{22}(t) X_2(t) = H_{21}(t) X_1(t) + W_2(t)$.

Toward this end, we have the following sequence of equalities,
\begin{eqnarray}
\lefteqn{ h \Big( Y_{2i}(t) \Big| Q(t), Y_{2[1:i-1]}(t) \Big) } \nonumber \\
&& {} \hspace{-1.2cm} =  h \Big( Y_{2i}(t) \Big| \mathcal{M}_2, \overline{H}(t), \overline{Y}_2(t-1), \overline{X}_2(t), Y_{2[1:i-1]}(t) \Big) \label{eq: lem: stst eqvalence step 1} \\
&& {} \hspace{-1.2cm} = h \Big( Y_{2i}'(t) \Big| \mathcal{M}_2, \overline{H}(t-1), \overline{Y}_2(t-1), \overline{X}_2(t), Y_{2[1:i-1]}'(t), H_{2[1:i]1}(t), H_{11}(t), H_{12}(t), H_{22}(t), H_{2[i+1:N_2]1}(t)  \Big) \label{eq: lem: stst eqvalence step 2} \\
&& {} \hspace{-1.2cm} = h \Big( Y_{2i}'(t) \Big| \mathcal{M}_2, \overline{H}(t-1), \overline{Y}_2(t-1), \overline{X}_2(t), Y_{2[1:i-1]}'(t), H_{2[1:i]1}(t) \Big) \label{eq: lem: stst eqvalence step 3} \\
&& {} \hspace{-1.2cm} = \mathbb{E}_{H_{2i1}(t) = a} ~ h \Big( Y_{2i}'(t) \Big| \mathcal{M}_2, \overline{H}(t-1), \overline{Y}_2(t-1), \overline{X}_2(t), Y_{2[1:i-1]}'(t), H_{2[1:i-1]1}(t), H_{2i1}(t) = a \Big) \label{eq: lem: stst eqvalence step 4} \\
&& {} \hspace{-1.2cm} = \mathbb{E}_{H_{2j1}(t) = a} ~ h \Big( Y_{2j}'(t) \Big| \mathcal{M}_2, \overline{H}(t-1), \overline{Y}_2(t-1), \overline{X}_2(t), Y_{2[1:i-1]}'(t), H_{2[1:i-1]1}(t), H_{2j1}(t) = a \Big) \label{eq: lem: stst eqvalence step 5} \\
&& {} \hspace{-1.2cm} = h \Big( Y_{2j}'(t) \Big| \mathcal{M}_2, \overline{H}(t-1), \overline{Y}_2(t-1), \overline{X}_2(t), Y_{2[1:i-1]}'(t), H_{2[1:i-1]1}(t), H_{2j1}(t) \Big) \label{eq: lem: stst eqvalence step 6} \\
&& {} \hspace{-1.2cm} = h \Big( Y_{2j}(t) \Big| \mathcal{M}_2, \overline{H}(t-1), \overline{Y}_2(t-1), \overline{X}_2(t), Y_{2[1:i-1]}(t), H(t) \Big) , \label{eq: lem: stst eqvalence step 7}
\end{eqnarray}
where the various equalities hold as follows:  (\ref{eq: lem: stst eqvalence step 1}) holds by the definition of $Q(t)$; (\ref{eq: lem: stst eqvalence step 2}) holds because translation does not change differential entropy \cite{CT}; (\ref{eq: lem: stst eqvalence step 3}) follows by noting that $\big\{ H_{11}(t), H_{12}(t), H_{22}(t), H_{2[i+1:N_2]1}(t)\big\}$ are independent of $\big\{ Y_{2i}'(t), \mathcal{M}_2, \overline{H}(t-1), \overline{Y}_2(t-1), \overline{X}_2(t), Y_{2[1:i-1]}'(t), H_{2[1:i]1}(t) \big\}$ (note the present channel matrices are independent of the present and the past channel inputs and noises); (\ref{eq: lem: stst eqvalence step 4}) holds by the definition of the conditional differential entropy; (\ref{eq: lem: stst eqvalence step 5}) holds because conditioned on $\big\{ \mathcal{M}_2, \overline{H}(t-1), \overline{Y}_2(t-1), \overline{X}_2(t), Y_{2[1:i-1]}'(t), H_{2[1:i-1]1}(t) \big\}$, the joint distribution of $\big\{ H_{2i1}(t), X_1(t), W_{2i}(t) \big\}$ is identical to that of $\big\{ H_{2j1}(t), X_1(t), W_{2j}(t) \big\}$; (\ref{eq: lem: stst eqvalence step 6}) holds by the definition of the conditional differential entropy;  (\ref{eq: lem: stst eqvalence step 7}) holds since $\big\{H_{2i1}(t), H_{2[i+2:N_2]1}(t), H_{11}(t), H_{12}(t), H_{22}(t)\big\}$ are independent of $\big\{ Y_{2j}'(t), \mathcal{M}_2, \overline{H}(t-1), \overline{Y}_2(t-1), \overline{X}_2(t), Y_{2[1:i-1]}'(t), H_{2[1:i-1]1}(t), H_{2j1}(t) \big\}$ and since translation does not change differential entropy.
\end{IEEEproof}

Note that the first equality in the above lemma asserts that the signals $Y_{2i}(t)$ and $Y_{2j}(t)$ received at the $i^{th}$ and $j^{th}$ antenna, respectively, of R2 have equal differential entropy, when conditioned on the channel matrices $\overline{H}(t)$, the message $\mathcal{M}_2$ and the transmit signal $\overline{X}_2(t)$ of T2, the past channel outputs $\overline{Y}_2(t-1)$, and the present channel outputs $Y_{2[1:i-1]}(t)$ at some other receive antennas. We refer to this property as the statistical equivalence of the channel outputs, which essentially says that given the past and present channel outputs, the signals received at any two antennas of the system provide an equal amount of information about $\mathcal{M}_1$. Note that this property of statistical equivalence of the channel outputs was shown to hold in \cite{Vaze_Varanasi_delay_MIMO_IC} for the case of delayed CSIT. Here, on the other hand, the same property is shown to be true under the stronger setting of Shannon feedback.

The above lemma yields the following simple corollary, where $Q(t) = \big\{ \mathcal{M}_2, \overline{H}(t), \overline{Y}_2(t-1), \overline{X}_2(t) \big\}$ as before.
\begin{corollary} %\label{lem: statistical equivalence dCSI+opfb 2user IC outer-bound}
For an $i \in [1:m_2-1]$ and a $k \in [1:m_1-m_2]$, if $j = i+1$ and $l = k+1$, we have the following:
\begin{eqnarray*}
h \Big( Y_{2i}(t) \Big| Q(t), Y_{2[1:i-1]}(t) \Big) & \geq & h \Big( Y_{2j}(t) \Big| Q(t), Y_{2[1:i]}(t) \Big); \\
h \Big( Y_{2m_2}(t) \Big| Q(t), Y_{2[1:m_2-1]}(t) \Big) & \geq & h \Big( Y_{11}(t) \Big| Q(t), Y_{2[1:m_2]}(t) \Big); \\
h \Big( Y_{1k}(t) \Big| Q(t), Y_{2[1:m_2]}(t), Y_{1[1:k-1]}(t) \Big) & \geq & h \Big( Y_{1l}(t) \Big| Q(t), Y_{2[1:m_2]}(t), Y_{1[1:k]}(t) \Big).
\end{eqnarray*}
\end{corollary}
\begin{IEEEproof}
Follows from the previous lemma by invoking the fact that conditioning reduces entropy \cite{CT}.
\end{IEEEproof}

%Consider next the following lemma.
\begin{lemma} \label{lem: penult lemma main ineq dCSI+opfb 2user IC}
We have
\[
m_1 \cdot h \Big( Y_{2[1:m_2]}(t) \Big| Q(t) \Big) \geq m_2 \cdot h \Big( Y_{1[1:m_1-m_2]}(t), Y_{2[1:m_2]}(t) \Big| Q(t), \overline{Y}_1(t-1) \Big).
\]
\end{lemma}
\begin{IEEEproof}
By the previous corollary and the chain rule for the differential entropy, we get
\begin{eqnarray*}
\lefteqn{ \frac{1}{m_2} h \Big( Y_{2[1:m_2]}(t) \Big| Q(t) \Big) = \frac{1}{m_2} \sum_{i=1}^{m_2} h \Big( Y_{2i}(t) \Big| Q(t), Y_{2[1:i-1]}(t) \Big) } \\
&& {} \geq  h \Big( Y_{2m_2}(t) \Big| Q(t), Y_{2[1:m_2-1]}(t) \Big) \\
&& {} \geq  h \Big( Y_{11}(t) \Big| Q(t), Y_{2[1:m_2]}(t) \Big) \\
&& {} \geq \frac{1}{m_1-m_2} h \Big( Y_{1[1:m_1-m_2]}(t) \Big| Q(t), Y_{2[1:m_2]}(t) \Big).
\end{eqnarray*}
This yields
\begin{eqnarray}
(m_1-m_2) \cdot h \Big( Y_{2[1:m_2]}(t) \Big| Q(t) \Big) & \geq & m_2 \cdot  h \Big( Y_{1[1:m_1-m_2]}(t) \Big| Q(t), Y_{2[1:m_2]}(t) \Big) \nonumber \\
& \geq & m_2 \cdot h \Big( Y_{1[1:m_1-m_2]}(t) \Big| Q(t), \overline{Y}_1(t-1), Y_{2[1:m_2]}(t) \Big) \label{eq: ineq1 penult lemma main ineq dCSI+dofb 2user IC}
\end{eqnarray}
since conditioning reduces entropy.
Similarly, we can obtain
\begin{eqnarray}
m_2 \cdot h \Big( Y_{2[1:m_2]}(t) \Big| Q(t) \Big) \geq  m_2 \cdot  h \Big( Y_{2[1:m_2]}(t) \Big| Q(t), \overline{Y}_1(t-1) \Big). \label{eq: ineq2 penult lemma main ineq dCSI+dofb 2user IC}
\end{eqnarray}
The lemma can now be obtained by adding the inequalities in (\ref{eq: ineq1 penult lemma main ineq dCSI+dofb 2user IC}) and (\ref{eq: ineq2 penult lemma main ineq dCSI+dofb 2user IC}).
\end{IEEEproof}

The inequality in Lemma \ref{lem: main inequality d-CSI+op-fb IC outer-bound} can now be derived by combining the results of Lemmas \ref{lem: extracting essential part of term1 main inequality dCSI+opfb IC outer-bound}, \ref{lem: extracting essential part of term2 main inequality dCSI+opfb IC outer-bound}, and \ref{lem: penult lemma main ineq dCSI+opfb 2user IC}, and by noting that the sum or the difference of two $o(\log_2 P)$ terms yields another $o(\log_2 P)$ term.

\section{Proof of Theorem \ref{thm: inner-bound 2user IC d-CSI and op fb}} \label{sec: proof of thm: inner-bound 2user IC d-CSI and op fb}

As mentioned before, it is sufficient to prove that the outer-bound $\mathbf{D}^{\rm S}_{\rm outer}$ is achievable when Condition $1$ holds. Throughout this section, it is assumed that Condition $1$ holds.

Here, bound $L_2$ can be easily shown to be redundant (it is implied by $L_3$), and thus can be ignored. Further, in the present case, bounds $L_1$ and $L_3$ are given by
\[
L_1 \equiv \frac{d_1}{M_1'} + \frac{d_2}{N_2} \leq 1 ~ \mbox{ and } ~ L_3 \equiv d_1 + d_2 \leq N_1,
\]
where $M_1' \define \min(M_1,N_1+N_2)$.

\begin{figure}[t] \centering
\includegraphics[bb=0bp 100bp 540bp 700bp,clip, height=3.5in, width=3.2in]{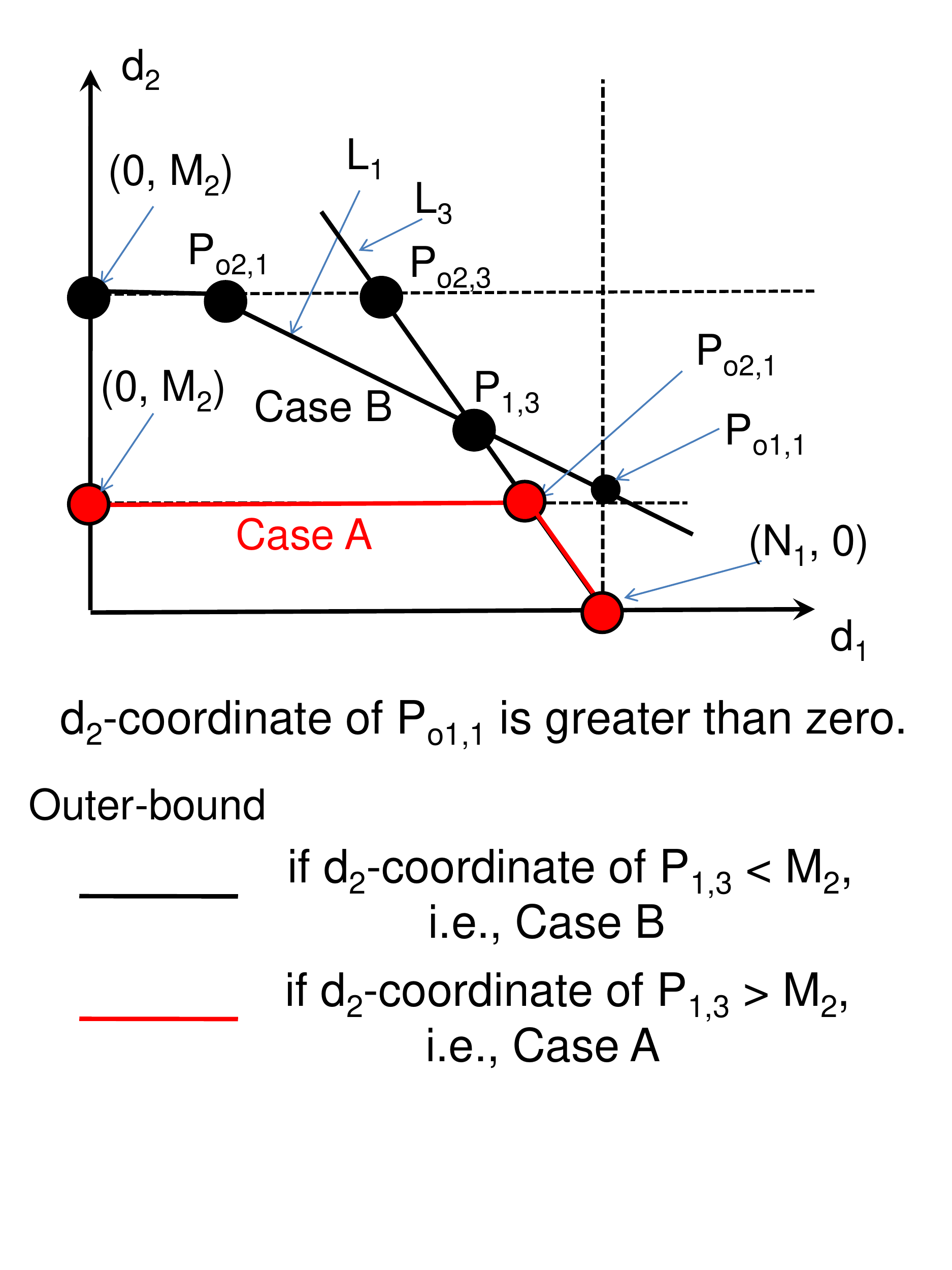}
\caption{Two Possible Shapes of the Outer-Bound when Condition $1$ Holds} \label{fig: Condition 1 dCSI+op-fb outerbound shape 2user IC}
\end{figure}

The typical shape of the outer-bound is shown in Fig. \ref{fig: Condition 1 dCSI+op-fb outerbound shape 2user IC}, where $P_{o2,1}$ is the point of intersection of the line $d_2 = M_2$ and the one corresponding to bound $L_1$, similarly $P_{o2,3}$, and $P_{1,3}$ is the point of intersection of lines corresponding to bounds $L_1$ and $L_3$. Moreover,
\[
P_{02,1} \equiv \left( M_1' \frac{N_2-M_2}{N_2}, M_2 \right), ~ \! P_{1,3} \equiv \left( M_1' \frac{N_1-N_2}{M_1'-N_2}, N_2 \frac{M_1'-N_1}{M_1'-N_2} \right), \mbox{ and } P_{o2,3} \equiv (N_1-M_2,M_2).
\]
Depending on whether the $d_2$-coordinate of $P_{1,3}$ is less than $M_2$ or not, we have to consider two cases separately. \newline $\bullet ~ \underline{\mbox{Case A: } M_1' \geq N_2 \frac{N_1-M_2}{N_2-M_2} ~ :}$

Here, bound $L_1$ is redundant. Moreover, from Fig. \ref{fig: Condition 1 dCSI+op-fb outerbound shape 2user IC}, one may observe that if $P_{o2,3} \in \mathbf{D}^{\rm S}$ then $\mathbf{D}^{\rm S}_{\rm outer} = \mathbf{D}^{\rm S}$. Hence, we find here sufficient to prove that $P_{o2,3} \in \mathbf{D}^{\rm S}$.  \newline $
\bullet ~ \underline{\mbox{Case B: } M_1' < N_2 \frac{N_1-M_2}{N_2-M_2} ~ :}$

Here, bounds $L_1$ and $L_3$ are both active. From Fig. \ref{fig: Condition 1 dCSI+op-fb outerbound shape 2user IC}, we observe the sufficiency of proving that $P_{o2,1}, P_{1,3} \in \mathbf{D}^{\rm S}$.

Next, we propose a generic retrospective interference alignment scheme, which is used later to prove that $P_{o2,3} \in \mathbf{D}^{\rm S}$ under Case A and $P_{o2,1}, P_{1,3} \in \mathbf{D}^{\rm S}$ under Case B with an appropriate choice of parameters. This scheme is specified in terms of the parameters
\begin{equation}
T, ~ t_1, ~ t_2, ~ \Big\{ m_1(i) \Big\}_{i=1}^{T}, ~ \mbox{ and } ~ \Big\{ m_2(i) \Big\}_{i=1}^{T}, \label{eq: parameters R-IA scheme Shannon journal}
\end{equation}
where $T, t_1, t_2 \in \mathbb{N}$, $m_1(i), m_2(i) \in \mathbb{N} \cup \{ 0 \}$ $ \forall$ $i$, and Design Criteria 1-5, which are stated in the sequel. It is developed such that if, for a given a DoF pair $P \equiv (d_1,d_2)$ and the given $(M_1,M_2,N_1,N_2)$ MIMO IC, the parameters in (\ref{eq: parameters R-IA scheme Shannon journal}) can be chosen as functions of $(d_1,d_2)$ and $(M_1,M_2,N_1,N_2)$ so that Design Criteria 1-5 are satisfied, then the DoF pair $P \equiv (d_1,d_2)\in \mathbf{D}^{\rm S}$ of the given $(M_1,M_2,N_1,N_2)$ MIMO IC.

%After describing this coding scheme, we will use it to shown that $P_{o2,3} \in \mathbf{D}^{\rm S}$ under Case A and $P_{o2,1}, P_{1,3} \in \mathbf{D}^{\rm S}$ under Case B \textcolor{red}{by appropriately choosing parameters in (\ref{eq: parameters R-IA scheme Shannon journal}).}

Consider now the retrospective interference alignment scheme. The goal is to prove that a given DoF pair $P \equiv (d_1,d_2) \in \mathbf{D}^{\rm S}$.
Let us first state two important design criteria.
\begin{design}
Choose positive integers $t_1$ and $t_2$ such that $t_1 + t_2 = T$.
\end{design}
\begin{design}
Choose a positive integer $T$ such that $T d_1$ and $T d_2$ are integers.
\end{design}

Now, choose a $B \in \mathbb{N}$ and set
\begin{eqnarray*}
T^{\star} = (B+1) \cdot T, ~ d_i^{\star} = B\cdot T \cdot d_i, \mbox{ where } i \in \{1,2\}, ~\mbox{ and } P^{\star} \equiv \left( \frac{d_1^{\star}}{T^{\star}}, \frac{d_1^{\star}}{T^{\star}} \right) = \left( \frac{d_1}{1 + \tfrac{1}{B}}, \frac{d_2}{1 + \tfrac{1}{B}} \right).
\end{eqnarray*}
It will be proved that for any positive integer $B$, by coding over $T^{\star}$ time slots, we can simultaneously achieve $d_1^{\star}$ and $d_2^{\star}$ DoF for the two users respectively. This implies that $P \in \mathbf{D}^{\rm S}$, since the DoF region is closed, and the point $P^{\star}$ converges to $P$ as $B \to \infty$. Thus, our aim in the following is to prove the achievability of point $P^{\star}$.

The entire duration of $T^{\star}$ is divided into $B+1$ blocks, each consisting of $T$ time slots. Each block is further divided into two phases with Phase One having $t_1$ time slots and Phase Two the remainder of $t_2 = T-t_1$ time slots. %It is convenient to be able to identify a time slot by the block to which it belongs and its index within that block.
\begin{definition}
Define two functions $b(t)$, the index of the block to which time slot $t$ belongs, and $\overline{t}(t)$,  the index of that time slot within Block $b(t)$, as
\[
b(t) = \left\lceil \frac{t}{T} \right\rceil \quad \mbox{ and } \quad \overline{t}(t) = t - T \cdot \big( b(t)-1 \big) .
\]
Note that $\overline{t}(t) \in [1:T]$. Thus, each time slot $t$ can be uniquely identified by the pair $\big( b(t), \overline{t}(t) \big)$. %, i.e., a time slot with $b(t)=b$ and $\overline{t}(t)=\overline{t}$ corresponds to $t=(b-1)T+\overline{t}$.
Block $b$, $b \in [1:B+1]$, consists of time slots $t \in [b'T + 1: b'T + T]$, where $b' \define b-1$; we let Phase One of block $b$ consist of time slots $t \in [b'T+1:b'T+t_1]$ and Phase Two the remaining time slots $t \in [b'T+t_1+1:b'T+t_1+t_2]$.
\end{definition}

The general structure of our achievability scheme has the following features:
\begin{itemize}
\item In each of the first $B$ blocks, T1 and T2 respectively transmit $T d_1$ and $T d_2$ i.i.d. complex Gaussian data symbols (DSs) intended for R1 and R2, respectively. In Block $B+1$, no new DS is sent.
\item In each time slot, T2 transmits an appropriate number of new DSs intended for R2, and therefore, in some sense, it does not play an active role in aligning interference (as in the example of Section  \ref{subsec: IA example dCSI_opfb 2user IC}).
\item T1, on the other hand, transmits all $T d_1$ DSs, to be transmitted over a given block during Phase One of that block (time slots $t=1$ or $t=4$ constitute Phase One in the example of Section \ref{subsec: IA example dCSI_opfb 2user IC}). It signals over Phase Two such that at the end of each block, (a) R2 can decode all DSs sent to it over that block, and (b) R1 can decode all DSs sent to it over the previous block (note that $t=2,3$ or $t=4,5$ are counterparts of Phase Two in the example of Section \ref{subsec: IA example dCSI_opfb 2user IC}). To meet these objectives optimally, T1 needs to align interference at both receivers.
\item Finally, the goal of Block $B+1$ is to allow R1 to decode all DSs of Block $B$ by not sending any new data (Block $B+1$, in some sense, corresponds to time slot $t=7$ in the example of Section    \ref{subsec: IA example dCSI_opfb 2user IC}).
\end{itemize}

Let T1 and T2 transmit $m_1(i)$ and $m_2(i)$ DSs intended for their respective receivers at the $i^{th}$ time slot of any given block (except, the last block) where these design parameters are chosen according to following criterion.
\begin{design}
Choose non-negative integers $m_1(i)$ and $m_2(i)$, $i \in [1:T]$ , as follows.
\begin{itemize}
\item $m_1(i) = 0$ $\forall$ $i \in [t_1+1:T]$ (recall, $t_1+t_2=T$);
\item $m_1(i) \leq M_1'$ $\forall$ $i \in [1:t_1]$ and $\sum_{i=1}^{T} m_1(i) = \sum_{i=1}^{t_1} m_1(i) = T d_1$;
\item $m_2(j) \leq M_2$ $\forall$ $j \in [1:T]$ and $\sum_{j=1}^{T} m_2(j) = T d_2$.
\end{itemize}
\end{design}
At time $t \in [1:BT]$, T1 transmits $m_1 \big( \overline{t}(t) \big)$ complex Gaussian DSs, denoted by $u_{1i}\big( b(t),\overline{t}(t) \big)$, whereas T2 transmits $m_2 \big( \overline{t}(t) \big)$ complex Gaussian DSs, denoted by $u_{2i}\big( b(t),\overline{t}(t) \big)$. Note that all DSs
\[
\left\{ \Big\{ \big\{ u_{1i}(b,\overline{t}) \big\}_{i=1}^{m_1(\overline{t})} \Big\}_{\overline{t}=1}^{T} \right\}_{b=1}^B ~ \mbox{and} ~ \left\{ \Big\{ \big\{ u_{2j}(b,\overline{t}) \big\}_{j=1}^{m_2(\overline{t})} \Big\}_{\overline{t}=1}^{T} \right\}_{b=1}^B
\]
are i.i.d. Note that since $m_1(i) = 0$ $\forall$ $i \in [t_1+1:t_1+t_2]$, no new DS is transmitted by T1 to R1 over Phase Two of any block so as to enable interference alignment and to ensure successful decoding.

The transmission scheme of T1 and T2 are described next. Focusing on Block $b$, where $b \in [1:B]$, the operation of T1 and T2 is described over the two phases separately. We start below with Phase One of Block $b$. See also Tables \ref{table: general scheme Block 1 dCSI_opfb 2user IC}-\ref{table: general scheme Block B+1 dCSI_opfb 2user IC}, where the main points about the operation of this scheme are summarized.

\underline{Block $b$, Phase One:} Here, $t \in [b'T+1:b'T+t_1]$ with $b \in [1:B]$ and $b' = b-1$. This phase is a data transmission phase.

At time $t \in [b'T+1:b'T+t_1]$ with $ b \in [1:B]$, T1 and T2 respectively transmit $m_1\big(\overline{t}(t)\big)$ and $m_2\big(\overline{t}(t)\big)$ DSs as follows:
\begin{eqnarray*}
X_{1i}(t) & = & u_{1i} \Big( b(t), \overline{t}(t) \Big), ~ \forall ~ \! i \in \Big[ 1:m_1 \Big( \overline{t}(t) \Big) \Big], \mbox{ and} \\
X_{1i}(t) & = & 0, ~ \forall ~ \! i \in \Big[ m_1 \Big( \overline{t}(t) \Big)+1 : M_1 \Big]; \mbox{ and} \\
X_{2j}(t) & = & u_{2j} \Big( b(t), \overline{t}(t) \Big), ~ \forall ~ \! j \in \Big [1 : m_2 \Big( \overline{t}(t) \Big) \Big], \mbox{ and} \\
X_{2j}(t) & = & 0, ~ \forall ~ \! j \in \Big[ m_2 \Big( \overline{t}(t) \Big)+1 : M_2 \Big].
\end{eqnarray*}
Consider now the signals received by R1 and R2 during Phase One. Since we are interested in the achievability of the DoF, we ignore throughout the presence of additive noise since it can not affect a DoF result. Then for a $t \in [b'T+1:b'T+t_1]$ with $ b \in [1:B]$, we have
\begin{eqnarray*}
Y_{1i}(t) & = & H_{1i1}(t) X_1(t) + H_{1i2}(t) X_2(t) ~ \cdots ~ i \in [1:N_1] \\
& = & \underbrace{ H_{1i1}(t) \begin{bmatrix} u_{11} \left( b(t), \overline{t}(t) \right) \\ u_{12} \left( b(t), \overline{t}(t) \right) \\ \vdots \\ u_{1 m_1 \left( \overline{t}(t) \right)} \left( b(t), \overline{t}(t) \right) \\ 0_{\left[ M_1 - m_1 \left( \overline{t}(t) \right) \right] \times 1} \end{bmatrix} }_{\define ~ {\rm LC}_{1i}^{[1]}\left(b(t),\overline{t}(t)\right) } + \underbrace{ H_{1i2}(t) \begin{bmatrix} u_{21} \left( b(t), \overline{t}(t) \right) \\ u_{22} \left( b(t), \overline{t}(t) \right) \\ \vdots \\ u_{2 m_2 \left( \overline{t}(t) \right)} \left( b(t), \overline{t}(t) \right) \\ 0_{\left[ M_2 - m_2 \left( \overline{t}(t) \right) \right] \times 1} \end{bmatrix} }_{\define ~ {\rm LC}_{1i}^{[2]}\left(b(t),\overline{t}(t)\right)}; \\
Y_{2j}(t) & = &  H_{2j1}(t) X_1(t) + H_{2j2}(t) X_2(t) ~ \cdots ~ j \in [1:N_2] \\
& = & \underbrace{ H_{2j1}(t) \begin{bmatrix} u_{11} \left( b(t), \overline{t}(t) \right) \\ u_{12} \left( b(t), \overline{t}(t) \right) \\ \vdots \\ u_{1 m_1 \left( \overline{t}(t) \right)} \left( b(t), \overline{t}(t) \right) \\ 0_{\left[ M_1 - m_1 \left( \overline{t}(t) \right) \right] \times 1} \end{bmatrix} }_{\define ~ {\rm LC}_{2j}^{[1]}\left(b(t),\overline{t}(t)\right)} + \underbrace{ H_{2j2}(t) \begin{bmatrix} u_{21} \left( b(t), \overline{t}(t) \right) \\ u_{22} \left( b(t), \overline{t}(t) \right) \\ \vdots \\ u_{2 m_2 \left( \overline{t}(t) \right)} \left( b(t), \overline{t}(t) \right) \\ 0_{\left[ M_2 - m_2 \left( \overline{t}(t) \right) \right] \times 1} \end{bmatrix} }_{\define ~ {\rm LC}_{2j}^{[2]}\left(b(t),\overline{t}(t)\right)}.
\end{eqnarray*}
Here, ${\rm LC}_{ij}^{[k]}\left(b,\overline{t}\right)$ represents the linear combination of DSs sent by the $k^{th}$ transmitter at time $t=(b-1)T+\overline{t}$, and it affects the signal received by the $i^{th}$ receiver at its $j^{th}$ antenna. Note here that ${\rm LC}_{2j}^{[1]}\left(b(t),\overline{t}(t)\right)$ is a linear combination of DSs intended only for R1, while it causes interference to R2. Note the collection
\[
\Big\{ \big\{ {\rm LC}^{[1]}_{2j}\left(b,\overline{t}\right) \big\}_{j=1}^{N_2} \Big\}_{\overline{t}=1}^{t_1}
\]
is referred to henceforth as the interference seen by R2 during Phase One of Block $b$. Evidently, interference at R2 is useful for R1. This completes the description of the operation over Phase One.

Consider next the second phase of Block $b \in [1:B]$. Here, T2 continues to transmit data. T1, on the other hand, employs two interference alignment techniques (described in Section \ref{subsec: intuition behind results dCSI_opfb 2user IC}) over Phase Two. In the first one, T1 transmits a part of the interference seen by R2 over Phase One of the previous block (this idea is not used during Block $1$), and under the second technique, it transmits all DSs of T2 sent earlier over Phase One of the same block. Now, as shown shortly, if over each time slot of Phase Two T1 transmits an appropriate numbers of DSs of T2 and the interfering linear combinations at R2 then at the end of each block R2 can decode the desired DSs sent over the same block, while R1 can decode those sent over the previous block.

{\renewcommand{\arraystretch}{1.6}
\begin{table}[t]
\begin{raggedright}
Phase One, Block 1: $t$ is such that $b(t)=1$ and $\overline{t}(t)\in[1:t_{1}]$.
\par\end{raggedright} \vspace{2mm}

\begin{centering}
\begin{tabular}{|c|c|}
\hline
node & operation at time t\tabularnewline
\hline
\hline
T1 & transmits DSs $u_{1i}(1,\overline{t}(t))$, $i\in[1:m_{1}(\overline{t}(t))]$.\tabularnewline
\hline
T2 & transmits DSs $u_{2j}(1,\overline{t}(t))$, $j\in[1:m_{2}(\overline{t}(t))]$.\tabularnewline
\hline
R1 & receives $Y_{1i}(t)={\rm LC}_{1i}^{[1]}(1,\overline{t}(t))+{\rm LC}_{1i}^{[2]}(1,\overline{t}(t))$,
$i\in[1:N_{1}]$. \tabularnewline
\hline
R2 & receives $Y_{2j}(t)={\rm LC}_{2j}^{[1]}(1,\overline{t}(t))+{\rm LC}_{2j}^{[2]}(1,\overline{t}(t))$,
$j\in[1:N_{2}]$. \tabularnewline
\hline
\end{tabular}
\par\end{centering}
\vspace{2mm}
\begin{raggedright}
Phase Two, Block 1: $t$ is such that $b(t)=1$ and $\overline{t}(t)\in[1:t_{2}]$.
\par\end{raggedright}
\vspace{2mm}
\begin{centering}
\begin{tabular}{|c|c|}
\hline
node & operation\tabularnewline
\hline
\hline
T1 & retransmits DSs $\left\{ \left\{ u_{2i}(1,\overline{t})\right\} _{i=1}^{m_{2}(\overline{t})}\right\} _{\overline{t}=1}^{t_{1}}$sent
by T2 over Phase One. \tabularnewline
 & At any time at most $N_{2}-M_{2}$ antennas are active.\tabularnewline
\hline
T2 & transmits $u_{2j}(1,\overline{t}(t))$, $j\in[1:m_{2}(\overline{t}(t))]$.\tabularnewline
\hline
R1 & Gets linear combinations $\left\{ \left\{ {\rm LC}_{1i}^{[1]}(1,\overline{t})\right\} _{i=1}^{N_{1}}\right\} _{\overline{t}=1}^{t_{1}}.$\tabularnewline
\hline
R2 & Can decode all DSs sent over this block. \tabularnewline
 & Knows the values of linear combinations $\left\{ \left\{ {\rm LC}_{2j}^{[1]}(1,\overline{t})\right\} _{j=1}^{N_{2}}\right\} _{\overline{t}=1}^{t_{1}}.$\tabularnewline
\hline
\end{tabular}
\par\end{centering}

\caption{Block $1$ of the retrospective interference alignment scheme with Shannon feedback} \label{table: general scheme Block 1 dCSI_opfb 2user IC}

\end{table} }

\begin{lemma} \label{lem: T1 knows T2's signal dCSI+opfb 2user IC}
At time $t = b'T + t_1+1$, T1 can obtain (a)  DSs $\Big\{ u_{2i} \Big( b(t),\overline{t}(t) \Big) \Big\}$ $\forall$ $i \in \Big[ 1:m_2 \Big( \overline{t} \big( t \big) \Big) \Big]$ and $t \in \Big[ b'T+1 : b'T+t_1 \Big]$, and (b) linear combinations $\Big\{ {\rm LC}_{2j}^{[1]} \Big( b(t), \overline{t}(t)\Big) \Big\}$, $\forall$ $j \in [1:N_2]$ and $t \in \Big[ b'T+1 : b'T+t_1 \Big]$.
\end{lemma}
\begin{IEEEproof}
Consider symbols in (a). By virtue of Shannon feedback, T1 knows all past channel matrices $H(t)$, the received signal $Y_2(t)$, as well as its own transmit signal. Hence, for each $t \in [b'T+1:b'T+t_1]$, it can compute
\[
Y_2(t) - H_{21}(t) X_1(t) = H_{22}(t) X_2(t).
\]
Since $N_2 > M_2$ under Condition $1$, and the channel matrices are Rayleigh faded, $H_{22}(t)$ is almost surely a one-to-one map. Hence, based on the value of $H_{22}(t) X_2(t)$, T1 can determine $X_2(t)$. In other words, at $t = b'T+t_1+1$, T1 can perfectly evaluate data symbols
\[
\Big\{ \big\{ u_{2i} \big( b(t), \overline{t} \big) \big\}_{i=1}^{m_2(\overline{t})} \Big\}_{\overline{t}=1}^{t_1}.
\]
That T1 can obtain the symbols in part (b) of the lemma can be shown analogously.
\end{IEEEproof}

%{\color{red} Thus, the two IA techniques are indeed feasible. meaning?}

Next, consider the transmission strategy of T1 over Phase Two. We need to construct two sets $\big\{ \mathcal{P}_{\rm LC} \big(b, i \big) \big\}_{i=1}^{t_2}$ and $\big\{ \mathcal{P}_{\rm DS} \big(b, i \big) \big\}_{i=1}^{t_2}$ for each $ b \in [1:B+1]$. The set $\mathcal{P}_{\rm LC} \big(b, i \big)$ contains the linear combinations that interfere with R2 over Phase One of Block $b-1$ (except, for $b=1$, for which this set is empty) and that are to be sent by T1 over the $i^{th}$ time slot of Phase Two of Block $b$. Moreover, the set $\mathcal{P}_{\rm DS} \big(b, i \big)$ contains DSs that are sent by T2 over Phase One of Block $b$ and that are to be retransmitted by T1 over the $i^{th}$ time slot of Phase Two of Block $b$. In terms of these sets the transmissions of T1 and T2 are then described.

%which is done in equations (\ref{eq: construct sets P_LC dCSI+opfb 2user IC}) and (\ref{eq: construct sets P_DS dCSI+opfb 2user IC}), respectively. Subsequently, the transmission policy of T1, and also of T2, over Phase Two is described. {\color{red} Here,  Note that the cardinality of these sets must be chosen appropriately so that the decoding is successful, and this is ensured by Design Criteria 4 and 5.}

With an aim of constructing sets $\Big\{ \mathcal{P}_{\rm LC} \big(b, i \big) \Big\}_{i=1}^{t_2}$ and $\Big\{ \mathcal{P}_{\rm DS} \big(b, i \big) \Big\}_{i=1}^{t_2}$ for each $b \in [1:B+1]$, consider the following. Let $(c-d)^+ \define \max\{0,c-d\}$, $|\mathcal{S}|$ denote the cardinality of set $\mathcal{S}$, $\mathcal{S}_{\rm LC}(1) = \phi$ (the empty set), and $\mathcal{S}_{\rm DS}(B+1) = \phi$. Further, define
\begin{eqnarray*}
n_{\rm req}(i) & \define & \Big( m_1(i) - N_1 \Big)^+, \mbox{ where } i \in [1:t_1]\\
\mbox{for a } b \in [2:B+1], \hspace{2pt} \mathcal{S}_{\rm LC}(b) & \define & \left\{ \Big\{ {\rm LC}_{2i}^{[1]} \left( b-1, \overline{t} \right) \Big\}_{i=1}^{n_{\rm req}(\overline{t})} \right\}_{\overline{t}=1}^{t_1}, \mbox{ and} \\
\mbox{for a } b \in [1:B], \hspace{2pt} \mathcal{S}_{\rm DS}(b) & \define & \left\{ \Big\{ u_{2i} \big(b,\overline{t})\Big\}_{i=1}^{m_2(\overline{t})} \right\}_{\overline{t}=1}^{t_1}.
\end{eqnarray*}
Here, the set $\mathcal{S}_{\rm LC}(b)$, $b \geq 2$, contains linear combinations that interfere with R2 over Block $b-1$ and the elements of this set are chosen such that if all linear combinations in this set are delivered to R1 then R1 can decode all desired DSs sent over Block $b-1$. Moreover, sets $\big\{ \mathcal{P}_{\rm LC} \big(b, i \big) \big\}_{i=1}^{t_2}$ are constructed by partitioning the set $\mathcal{S}_{\rm LC}(b)$. Further, the set $\mathcal{S}_{\rm DS}(b)$, $b \leq B$, contains all DSs sent by T2 over Phase One of Block $b$, and by partitioning this set, smaller sets $\Big\{ \mathcal{P}_{\rm DS} \big(b, i \big) \Big\}_{i=1}^{t_2}$ are formed.

Note that
\[
\Big| \mathcal{S}_{\rm LC}(b) \Big| = \sum_{\overline{t}=1}^{t_1} n_{\rm req}(\overline{t}) \mbox{ and } \Big| \mathcal{S}_{\rm DS}(b) \Big| = \sum_{\overline{t}=1}^{t_1} m_2(\overline{t}).
\]
Consider next two more design criteria which ensure that the cardinalities of these sets are appropriately bounded.
\begin{design}
Choose $t_1$, $t_2$, and $m_2(\overline{t})$, where $\overline{t} \in [1:t_1]$, such that
\[
\Big| \mathcal{S}_{\rm DS}(b) \Big| = \sum_{\overline{t}=1}^{t_1} m_2(\overline{t}) \leq (N_2-M_2) \cdot t_2.
\]
\end{design}
\begin{design}
Choose $t_1$, $t_2$, and $m_1(\overline{t})$, where $\overline{t} \in [1:t_1]$, such that
\[
\Big| \mathcal{S}_{\rm LC}(b) \Big| = \sum_{\overline{t}=1}^{t_1} n_{\rm req}(\overline{t}) \leq (N_1-N_2) \cdot t_2.
\]
\end{design}
If these two criteria are satisfied, sets $\mathcal{S}_{\rm DS}(b)$ and $\mathcal{S}_{\rm LC}(b)$ can be partitioned as follows.
\begin{itemize}
\item Partition $\mathcal{S}_{\rm LC}(b)$ into $t_2$ disjoint subsets each of cardinality at most $N_1-N_2$ so that
    \begin{equation}
    \mathcal{S}_{\rm LC}(b) = \bigcup_{i=1}^{t_2} \mathcal{P}_{\rm LC} \big(b, i \big) \mbox{ and } \left| \mathcal{P}_{\rm LC} \big(b, i \big) \right| \leq N_1-N_2. \label{eq: construct sets P_LC dCSI+opfb 2user IC}
    \end{equation}
\item Partition $\mathcal{S}_{\rm DS}(b)$ into $t_2$ disjoint subsets each of cardinality at most $N_2 - M_2$ so that
    \begin{equation}
    \mathcal{S}_{\rm DS}(b) = \bigcup_{i=1}^{t_2}  \mathcal{P}_{\rm DS} \big(b, i \big) \mbox{ and } \left| \mathcal{P}_{\rm DS} \big(b, i \big) \right| \leq N_2-M_2. \label{eq: construct sets P_DS dCSI+opfb 2user IC}
    \end{equation}
\end{itemize}
Suppose for each $i \in [1:t_2]$,
\[
\mathcal{P}_{\rm LC} \big(b, i \big) \bigcup \mathcal{P}_{\rm DS} \big(b, i\big) = \Big\{ p_j \big( b,i \big) \Big\}, ~ j \in \Big[ 1 : \left| \mathcal{P}_{\rm LC} \big(b, i \big) \right| + \left| \mathcal{P}_{\rm DS} \big(b, i \big) \right| \Big].
\]

{\renewcommand{\arraystretch}{1.6}
\begin{table}[t]
\begin{raggedright}
Phase One, Block $b$, $b\in[2:B]$:
\end{raggedright}

\vspace{2mm}
\begin{centering}
\begin{tabular}{|c|c|}
\hline
node & operation at time t\tabularnewline
\hline
\hline
T1 & transmits DSs $u_{1i}(b,\overline{t}(t))$, $i\in[1:m_{1}(\overline{t}(t))]$.\tabularnewline
\hline
T2 & transmits DSs $u_{2j}(b,\overline{t}(t))$, $j\in[1:m_{2}(\overline{t}(t))]$.\tabularnewline
\hline
R1 & receives $Y_{1i}(t)={\rm LC}_{1i}^{[1]}(b,\overline{t}(t))+{\rm LC}_{1i}^{[2]}(b,\overline{t}(t))$,
$i\in[1:N_{1}]$. \tabularnewline
\hline
R2 & receives $Y_{2j}(t)={\rm LC}_{2j}^{[1]}(b,\overline{t}(t))+{\rm LC}_{2j}^{[2]}(b,\overline{t}(t))$,
$j\in[1:N_{2}]$. \tabularnewline
\hline
\end{tabular}
\par\end{centering}
\vspace{2mm}
\begin{raggedright}
Phase Two, Block $b$, $b\in[2:B]$:
\end{raggedright}
\vspace{2mm}

\begin{centering}
\begin{tabular}{|c|c|}
\hline
node & operation\tabularnewline
\hline
\hline
T1 &  transmits DSs $\left\{ \left\{ u_{2i}(b,\overline{t})\right\} _{i=1}^{m_{2}(\overline{t})}\right\} _{\overline{t}=1}^{t_{1}}$and
linear combinations $\left\{ \left\{ {\rm LC}_{2j}^{[1]}\left(b-1,\overline{t}\right)\right\} _{j=1}^{n_{{\rm req}}(\overline{t})}\right\} _{\overline{t}=1}^{t_{1}}$. \tabularnewline
 & At any time, at most $N_{2}-M_{2}$ DSs and $N_{1}-N_{2}$ linear combinations
are transmitted.\tabularnewline
\hline
T2 & transmits $u_{2j}(b,\overline{t}(t))$, $j\in[1:m_{2}(\overline{t}(t))]$.\tabularnewline
\hline
R1 & Gets linear combinations $\left\{ \left\{ {\rm LC}_{1i}^{[1]}(b,\overline{t})\right\} _{i=1}^{N_{1}}\right\} _{\overline{t}=1}^{t_{1}}$and
$\left\{ \left\{ {\rm LC}_{2j}^{[1]}\left(b-1,\overline{t}\right)\right\} _{j=1}^{n_{{\rm req}}(\overline{t})}\right\} _{\overline{t}=1}^{t_{1}}$. \tabularnewline
 & Can decode all DSs sent over Block $b-1$.\tabularnewline
\hline
R2 & Can decode all DSs sent over this Block. \tabularnewline
 & Knows the values of linear combinations $\left\{ \left\{ {\rm LC}_{2j}^{[1]}(b,\overline{t})\right\} _{j=1}^{N_{2}}\right\} _{\overline{t}=1}^{t_{1}}.$\tabularnewline
\hline
\end{tabular}
\par\end{centering}

\caption{Block $b$, $b\in[2:B]$, of retrospective IA scheme with Shannon feedback} \label{table: general scheme Block b dCSI_opfb 2user IC}

\end{table} }

Over Phase Two of Block $b$, T1 transmits all the elements of the set $\mathcal{S}_{\rm LC}(b) \bigcup \mathcal{S}_{\rm DS}(b)$, while T2 continues to transmit DSs intended for R2. See also Tables \ref{table: general scheme Block 1 dCSI_opfb 2user IC}-\ref{table: general scheme Block B+1 dCSI_opfb 2user IC}.

\underline{Block $b$, Phase Two:} Here, $t \in [b'T+t_1+1:b'T+t_1+t_2]$.

At time $t \in [b'T+t_1+1:b'T+t_1+t_2]$ with $b \in [1:B+1]$, T1 transmits the elements of set $\mathcal{P}_{\rm LC} \Big(b, \overline{t}(t)-t_1  \Big) \bigcup \mathcal{P}_{\rm DS} \Big(b, \overline{t}(t)-t_1  \Big)$ as follows:
\begin{eqnarray*}
X_{1i}(t) = \begin{cases} p_i \Big( b(t), \overline{t}(t)-t_1 \Big), & \mbox{if } i \in \Big[ 1 : \left| \mathcal{P}_{\rm LC} \Big( b(t), \overline{t}(t)-t_1 \Big) \right| + \left| \mathcal{P}_{\rm DS} \Big( b(t), \overline{t}(t)-t_1 \Big) \right| \Big], \\
0, & \mbox{if } i \in \Big[ 1 + \left| \mathcal{P}_{\rm LC} \Big( b(t), \overline{t}(t)-t_1 \Big) \right| + \left| \mathcal{P}_{\rm DS} \Big( b(t), \overline{t}(t)-t_1 \Big) \right| : M_1 \Big]. \end{cases}
\end{eqnarray*}

During Phase Two of Block $b$, T2 transmits DSs $\Big\{ \big\{ u_{2j}(b,\overline{t}) \big\}_{j=1}^{m_2(\overline{t})} \Big\}_{\overline{t}=t_1+1}^{T}$ if $b \leq B$, and remains silent over the last Block. Thus, we have the following: for $t \in [b'T+t_1+1:b'T+t_1+t_2]$ with $b \leq B$,
\begin{eqnarray*}
X_{2j}(t) & = & u_{2j} \Big( b(t), \overline{t}(t) \Big), ~ j \in \Big[ 1:m_2 \Big( \overline{t}(t) \Big) \Big], \mbox{ and} \\
X_{2j}(t) & = & 0, ~ j \in \Big[ m_2 \Big( \overline{t}(t) \Big)+1:M_2 \Big],
\end{eqnarray*}
whereas for $t \in [BT+t_1+1:BT+t_1+t_2]$ (i.e., over Block $B+1$),
\[
X_2(t) = 0.
\]

This completes the description of the transmission strategy of T1 and T2.

Consider now the decoding operation starting with R2. The following two lemmas enable an inductive proof that R2 can decode all desired DSs. See also Tables \ref{table: general scheme Block 1 dCSI_opfb 2user IC}-\ref{table: general scheme Block B+1 dCSI_opfb 2user IC}.

The next lemma proves that R2 can decode all DSs sent to it over Block $1$.
\begin{lemma}
At time $t=T$, R2 can decode DSs
\[
\Big\{ \big\{ u_{2j}(1,\overline{t}) \big\}_{j=1}^{m_2(\overline{t})} \Big\}_{\overline{t}=1}^{T}
\]
sent to it over Block $1$.
\end{lemma}
\begin{IEEEproof}
Recall that for Block $1$, the set $\mathcal{S}_{\rm LC}(1)$ is empty. Thus, over Phase Two of Block $1$, T1 transmits all elements of the set $\mathcal{S}_{\rm DS}(1)$, i.e., T1 retransmits all DSs $\left\{ \big\{ u_{2i}(1,\overline{t}) \big\}_{i=1}^{m_2(\overline{t})} \right\}_{\overline{t}=1}^{t_1} $ that are sent by T2 over Phase One of this block. Moreover, since $\left| \mathcal{P}_{\rm DS} \left( 1, i \right) \right| \leq N_2-M_2$, $\forall$ $i,b$, T1, at any time during Phase Two, transmits at most $N_2-M_2$ elements of the set $\mathcal{S}_{\rm DS}(1) = \left\{ \big\{ u_{2i}(1,\overline{t}) \big\}_{i=1}^{m_2(\overline{t})} \right\}_{\overline{t}=1}^{t_1}$. This implies that at any time during Phase Two of Block $1$, at most $N_2$ transmit antennas are active (i.e., they send a non-zero signal). Since the Rayleigh-faded channel matrices are full rank almost surely, R2, equipped with $N_2$ antennas, can determine the transmit signal via simple channel inversion. Therefore, at time $t \in [t_1+1:T]$, R2 can decode all DSs belonging to the set $\mathcal{P}_{\rm DS} \Big( b(t), \overline{t}(t)-t_1 \Big)$ and also those transmitted by T2 at that time, namely, $u_{2i} \Big( 1,\overline{t}(t) \Big)$, $i \in \Big[ 1: m_2 \big( \overline{t}(t) \big) \Big]$. Therefore, at the end of Block $1$, i.e., at $t=T$, R2 can decode all DSs sent to it over this block.
\end{IEEEproof}

The next lemma deals with decoding of DSs sent to R2 over Block $b$, $b \in [2:B]$.

\begin{lemma}
Consider Block $b$, $b \in [2:B]$. If, at time $t = (b-1)T$, R2 has successfully decoded all DSs
\[
\left\{ \Big\{ u_{2i}(b-1,\overline{t}) \Big\}_{i=1}^{m_2(\overline{t})} \right\}_{\overline{t}=1}^T
\]
sent to it over Block $(b-1)$, then at time $t = bT$, it can decode all DSs
\[
\left\{ \Big\{ u_{2i}(b,\overline{t}) \Big\}_{i=1}^{m_2(\overline{t})} \right\}_{\overline{t}=1}^T
\]
sent to it over Block $b$.
\end{lemma}
\begin{IEEEproof}
Suppose at time $t = (b-1) T$, R2 has decoded successfully all DSs sent to it over Block $(b-1)$. Thus, R2, at time $t = (b-1) T$, can determine the values of linear combinations
\[
{\rm LC}_{2j}^{[2]}\left(b-1,\overline{t}\right) = H_{2j2}\Big( (b-2)T+\overline{t} \Big) \begin{bmatrix} u_{21} \left( b-1, \overline{t} \right) \\ u_{22} \left( b-1, \overline{t} \right) \\ \vdots \\ u_{2 m_2( \overline{t})} \left( b-1, \overline{t} \right) \\ 0_{\left[ M_2 - m_2( \overline{t}) \right] \times 1} \end{bmatrix}, ~ \forall ~ \! j \in [1:N_2] \mbox{ and } \overline{t} \in [1:t_1],
\]
and hence, it can also evaluate
\[
{\rm LC}_{2j}^{[1]}\left(b-1,\overline{t}\right) = Y_{2j}\Big( (b-2)T+\overline{t} \Big) - {\rm LC}_{2j}^{[2]}\left(b-1,\overline{t}\right), ~ \forall ~ \! j \in [1:N_2] \mbox{ and } \overline{t} \in [1:t_1].
\]
In particular, R2 at time $t = (b-1)T$ knows values of all elements of set
\[
\mathcal{S}_{\rm LC}(b) = \left\{ \Big\{ {\rm LC}_{2i}^{[1]} \left( b-1, \overline{t} \right) \Big\}_{i=1}^{n_{\rm req}(\overline{t})} \right\}_{\overline{t}=1}^{t_1}.
\]

Consider now the operation over Phase Two of Block $b$, i.e., for a $t \in [(b-1)T+t_1+1:(b-1)T+t_1+t_2]$. Since R2 already knows the values of elements of set $\mathcal{S}_{\rm LC}(b)$, it can subtract from $Y_2(t)$ the contribution due to the elements of set $\mathcal{P}_{\rm LC} \big(b, \overline{t}(t)-t_1 \big) \subset \mathcal{S}_{\rm LC}(b)$, which are transmitted by T1 at time $t$. After this subtraction, from the perspective of R2, not more than $N_2$ transmit antennas are active. Therefore, as mentioned before, R2 can use channel inversion to determine the elements of set $\mathcal{P}_{\rm DS} \big( b,\overline{t}(t)-t_1 \big)$ and also $u_{2i}\big( b,\overline{t}(t) \big)$ $\forall$ $i \in [1:m_2(\overline{t}(t))]$.
\end{IEEEproof}

Combined with the result of the previous lemma that all Block $1$ DSs can be decided at $t = T$, the above lemma can now be applied recursively to show that R2 can decode all desired DSs sent over the first $B$ blocks. Since no new DSs are transmitted over the last $(B+1)^{th}$ block, decoding is successful at R2.

%Using these two lemmas, we can easily see that at the end of Block $B$, i.e., at $t = BT$, R2 can decode all desired DSs.

%
{\renewcommand{\arraystretch}{1.5}
\begin{table}[t]
\begin{raggedright}
Phase One, Block $B+1$: No operation is performed.
\end{raggedright}

\vspace{2mm}
\begin{raggedright}
Phase Two, Block $B+1$:
\end{raggedright}

\vspace{2mm}
\begin{centering}
\begin{tabular}{|c|c|}
\hline
node & operation\tabularnewline
\hline
\hline
T1 & transmits linear combinations $\left\{ \left\{ {\rm LC}_{2j}^{[1]}\left(B,\overline{t}\right)\right\} _{j=1}^{n_{{\rm req}}(\overline{t})}\right\} _{\overline{t}=1}^{t_{1}}$. \tabularnewline
\hline
T2 & remains silent\tabularnewline
\hline
R1 & Gets linear combinations$\left\{ \left\{ {\rm LC}_{2j}^{[1]}\left(B,\overline{t}\right)\right\} _{j=1}^{n_{{\rm req}}(\overline{t})}\right\} _{\overline{t}=1}^{t_{1}}$. \tabularnewline
 & Can decode all DSs sent over Block $B$.\tabularnewline
\hline
R2 & idle\tabularnewline
\hline
\end{tabular}
\par\end{centering}

\caption{Block $B+1$ of retrospective IA scheme with Shannon feedback} \label{table: general scheme Block B+1 dCSI_opfb 2user IC}

\end{table}
}

Consider now the decoding procedure at R1. See also Tables \ref{table: general scheme Block 1 dCSI_opfb 2user IC}-\ref{table: general scheme Block B+1 dCSI_opfb 2user IC}. It turns out that R1, at the end of a given block, does not observe a sufficient number of interference-free linear combinations required to decode desired DSs sent over that block. However, the missing linear combinations are sent to it over the next block. Hence, DSs sent over a given block are decodable at R1 at the end of the next block. Since no new DS is sent over the final block, decoding is successful at R1 at the end of Block $B+1$.

The above claims about how decoding works at R1 are proved using a series of three lemmas. The first two specify the linear combinations that are known to R1 at the end of each block. The third lemma makes use of the first two to prove that all desired DSs are decodable at R1 at the end of Block $B+1$.

\begin{lemma}
At the end of Block $1$, i.e., time $t=T$, R1 can obtain linear combinations
\[
{\rm LC}_{1i}^{[1]} \left( 1,\overline{t}(t) \right) ~ \forall ~\! i \in [1:N_1] \mbox{ and } t \in [1:t_1].
\]
\end{lemma}
\begin{IEEEproof}
As with R2, R1 at $t=T$ can determine the DSs
\[
\Big\{ \big\{ u_{2i}\big( 1,\overline{t}(t) \big) \big\}_{i=1}^{m_2(\overline{t}(t))} \Big\}_{t=1}^{t_1} .
\]
Hence it can evaluate
\[
{\rm LC}_{1i}^{[2]} \left( 1,\overline{t} \right) = H_{1i2}(\overline{t}) \begin{bmatrix} u_{21} \left( 1, \overline{t} \right) \\ u_{22} \left( 1, \overline{t} \right) \\ \vdots \\ u_{2 m_2( \overline{t})} \left( 1, \overline{t} \right) \\ 0_{\left[ M_2 - m_2( \overline{t}) \right] \times 1} \end{bmatrix}, ~ \forall ~ \! i \in [1:N_1] \mbox{ and } \overline{t} \in [1:t_1],
\]
and then, $ {\rm LC}_{1i}^{[1]} \left( 1,\overline{t} \right) = Y_{1i}(\overline{t}) - {\rm LC}_{1i}^{[2]} \left( 1,\overline{t} \right), ~ \forall ~ \! i \in [1:N_1] \mbox{ and } \overline{t} \in [1:t_1].$
\end{IEEEproof}

\begin{lemma}
At the end of Block $b$, $b \in [2:B]$, i.e., at $t = bT$, R1 can obtain the linear combinations \[
\Big\{ \big\{ {\rm LC}_{1i}^{[1]} \big( b,\overline{t} \big) \big\}_{i=1}^{N_1} \Big\}_{\overline{t}=1}^{t_1} \mbox{ and } \Big\{ \big\{ {\rm LC}_{2i}^{[1]} \big( b-1,\overline{t} \big) \big\}_{i=1}^{n_{\rm req}(\overline{t})} \Big\}_{\overline{t}=1}^{t_1}.
\]
Further, at the end of Block $B+1$, R1 can obtain the linear combinations
\[
\Big\{ \big\{ {\rm LC}_{2i}^{[1]} \big( B,\overline{t} \big) \big\}_{i=1}^{n_{\rm req}(\overline{t})} \Big\}_{\overline{t}=1}^{t_1}.
\]
\end{lemma}
\begin{IEEEproof}
Consider the operation over Phase Two of Block $b$, $b \leq B$ i.e., for $t \in [b'T+t_1+1 : b'T+t_1+t_2] $ with $b \leq B$. At any time during this phase, at most $N_1$ transmit antennas are active. Hence, via simple channel inversion, R1 can determine the transmit signal during this phase. Thus, at the end of this phase, i.e., at $t = bT$, R1 knows the values of elements of the sets $\mathcal{S}_{\rm LC}(b)$ and $\mathcal{S}_{\rm DS}(b)$. This implies that at the end of Block $b$ R1 knows linear combinations $\Big\{ \big\{ {\rm LC}_{2i}^{[1]} \big( b-1,\overline{t} \big) \big\}_{i=1}^{n_{\rm req}(\overline{t})} \Big\}_{\overline{t}=1}^{t_1}$, which are contained in the set $\mathcal{S}_{\rm LC}(b)$. Further, it knows DSs $\Big\{ \big\{ u_{2i}(b,\overline{t}) \big\}_{i=1}^{m_2(\overline{t})} \Big\}_{\overline{t}=1}^{t_1}$, from which it can compute $\Big\{ \big\{ {\rm LC}_{1i}^{[2]} \big( b,\overline{t} \big) \big\}_{i=1}^{N_1} \Big\}_{\overline{t}=1}^{t_1}$, and subsequently obtain $\Big\{ \big\{ {\rm LC}_{1i}^{[1]} \big( b,\overline{t} \big) \big\}_{i=1}^{N_1} \Big\}_{\overline{t}=1}^{t_1}$.

The last part of the lemma can be proved using similar arguments.
\end{IEEEproof}

\begin{lemma}
At the end of Block $b$, where $b \in [2:B+1]$, R1 can decode data symbols
\[
\Big\{ \big\{ u_{1i}(b-1,\overline{t}) \big\}_{i=1}^{m_1(\overline{t})} \Big\}_{\overline{t}=1}^{t_1}
\]
sent to it over Block $b-1$.
\end{lemma}
\begin{IEEEproof}
First recall that no new DSs are transmitted to R1 during Block $B+1$. Therefore, we will focus on Block $b-1$, where $b \in [2:B+1]$. It will be shown that given any $t$ such that $b(t) = b-1$, all DSs sent by T1 to R1 at time $t$ can be decoded by R1 at the end of Block $b$. Moreover, since no new DSs are sent to R1 over Phase Two of any block, we may assume, without loss of generality, that $\overline{t}(t) \in [1:t_1]$. Thus, in the following, we consider a time slot $t$ with $b(t) = (b-1)$, $ b \in [2:B+1]$, and $\overline{t}(t) \in [1:t_1]$ or $t = (b-2)T+\overline{t}$ with $\overline{t} \in [1:t_1]$.

At this time, T1 transmits DSs
\[
\Big\{ u_{1i} \Big( b-1,\overline{t}(t) \Big) \Big\}, i \in \Big[ 1: m_1 \big( \overline{t}(t) \big) \Big].
\]
To decode these DSs, it is sufficient for R1 to know $m_1 \Big( \overline{t} \big( t \big) \Big)$ linearly independent linear combinations of these data symbols. Consider the following collection of linear combinations:
\[
\Big\{ {\rm LC}_{1i}^{[1]} \big( b-1,\overline{t}(t) \big) \Big\}_{i=1}^{N_1} \mbox{ and } \Big\{ {\rm LC}_{2i}^{[1]} \big( b-1,\overline{t}(t) \big) \Big\}_{i=1}^{n_{\rm req} \left( \overline{t}(t) \right)}.
\]
These are $m_1 \big( \overline{t}(t) \big)$ linear combinations of DSs $\big\{ u_{1i} \big( b-1,\overline{t}(t) \big) \big\}$, $i \in \left[ 1: m_1 \big( \overline{t}(t) \big) \right]$, and since the Rayleigh-faded channel matrices are full rank with probability $1$, these linear combinations are almost surely linearly independent. Hence, if R1 knows the values of these linear combinations, it can decode data symbols transmitted at time $t$.

Moreover, by combining the results of previous two lemmas, we observe that R1 can determine the linear combinations $
\Big\{ {\rm LC}_{1i}^{[1]} \big( b-1,\overline{t}(t) \big) \Big\}_{i=1}^{N_1} $ at the end of Block $(b-1)$, while it can obtain linear combinations
$ \Big\{ {\rm LC}_{2i}^{[1]} \big( b-1,\overline{t}(t) \big) \Big\}_{i=1}^{n_{\rm req} \left( \overline{t}(t) \right)} $
at the end of Block $b$. Thus, at the end of Block $b$, R1 can decode all DSs sent to it at time $t$, which belongs to Phase One of Block $b-1$. Hence, at the end of Block $b$, R1 can decode all DSs sent to it over Block $b-1$.
\end{IEEEproof}

Thus, we conclude that by coding over $T^{\star}$ time slots, $(d_1^{\star}, d_2^{\star})$ DoF can be achieved as desired. This completes the description of our generic retrospective interference alignment scheme.

We will now use this scheme to prove that $P_{o2,3} \in \mathbf{D}^{\rm S}$ under Case A and
$P_{o2,1}, P_{1,3} \in \mathbf{D}^{\rm S}$ under Case B in the following three subsections.

\subsection{Proof of $P_{o2,3} \in \mathbf{D}^{\rm S}$ under Case A}
Recall that under Case A, Condition $1$ holds and
\[
M_1' \geq N_2 \frac{N_1-M_2}{N_2 - M_2},
\]
and $P_{o2,3} \equiv (d_1,d_2) =(N_1-M_2,M_2)$.

We use the generic retrospective interference alignment scheme with the parameters chosen as follows:
\[
T = N_2, ~ t_1 = N_2 - M_2, ~ t_2 = M_2, \mbox{ and } m_2(\overline{t}) = M_2 ~ \forall ~ \! \overline{t} \in [1:T].
\]
It is easy to verify that this choice satisfies Design Criteria 1 and 2.

In order to  choose $m_1(\overline{t})$, $\overline{t} \in [1:t_1]$, consider the following:
\begin{eqnarray*}
\left\lceil N_2 \frac{N_1-M_2}{N_2 - M_2} \right\rceil & \geq & N_2 \frac{N_1-M_2}{N_2 - M_2} \geq \left\lfloor N_2 \frac{N_1-M_2}{N_2 - M_2} \right\rfloor \\
\Rightarrow t_1 \cdot \left\lceil N_2 \frac{N_1-M_2}{N_2 - M_2} \right\rceil & \geq &  t_1 \cdot N_2 \frac{N_1-M_2}{N_2 - M_2} \geq t_1 \cdot \left\lfloor N_2 \frac{N_1-M_2}{N_2 - M_2} \right\rfloor \\
\Rightarrow  t_1 \cdot \left\lceil N_2 \frac{N_1-M_2}{N_2 - M_2} \right\rceil & \geq & N_2(N_1-M_2) = T d_1 \geq t_1 \cdot \left\lfloor N_2 \frac{N_1-M_2}{N_2 - M_2} \right\rfloor,
\end{eqnarray*}
This suggests that we can choose $m_1(\overline{t})$, $\overline{t} \in [1:t_1]$ as
\begin{eqnarray*}
m_1(\overline{t}) & \in & \left\{ \left\lfloor N_2 \frac{N_1-M_2}{N_2 - M_2} \right\rfloor, \left\lceil N_2 \frac{N_1-M_2}{N_2 - M_2} \right\rceil \right\}, ~ \forall \overline{t} \in [1:t_1], \mbox{ such that } \nonumber \\
T d_1 & = & N_2 (N_1-M_2) = \sum_{\overline{t}=1}^{t_1} m_1(\overline{t}), %\label{eq: ia-retro-coop m1eff choice dCSI+opfb 2user IC} \\
\end{eqnarray*}
where $\lfloor x \rfloor$ denotes the largest integer that is less than or equal to $x$. It can be verified that this choice of parameters satisfies Design Criterion 3. Furthermore, Design Criterion 4 holds since $(N_2-M_2) t_2 = (N_2-M_2) M_2 $ and $\sum_{\overline{t}=1}^{t_1} m_2(\overline{t}) = M_2 t_1 = M_2 (N_2-M_2)$. Next we prove that Design Criteria 5 is satisfied as well.

%\underline{Proof that Design Criterion 3 holds:}
%
%First note that since $M_1 \geq N_2 \frac{N_1-M_2}{N_2 - M_2}$, we have
%\[
%M_1' \geq \left\lfloor N_2 \frac{N_1-M_2}{N_2 - M_2} \right\rfloor, \left\lceil N_2 \frac{N_1-M_2}{N_2 - M_2} \right\rceil,
%\]
%which implies that $m_1(\overline{t}) \leq M_1'$ $\forall$ $\overline{t} \in [1:t_1]$ as needed. Consider now the following:
%\begin{eqnarray*}
%\left\lceil N_2 \frac{N_1-M_2}{N_2 - M_2} \right\rceil & \geq & N_2 \frac{N_1-M_2}{N_2 - M_2} \geq \left\lfloor N_2 \frac{N_1-M_2}{N_2 - M_2} \right\rfloor \\
%\Rightarrow t_1 \cdot \left\lceil N_2 \frac{N_1-M_2}{N_2 - M_2} \right\rceil & \geq &  t_1 \cdot N_2 \frac{N_1-M_2}{N_2 - M_2} \geq t_1 \cdot \left\lfloor N_2 \frac{N_1-M_2}{N_2 - M_2} \right\rfloor \\
%\Rightarrow  t_1 \cdot \left\lceil N_2 \frac{N_1-M_2}{N_2 - M_2} \right\rceil & \geq & N_2(N_1-M_2) = T d_1 \geq t_1 \cdot \left\lfloor N_2 \frac{N_1-M_2}{N_2 - M_2} \right\rfloor,
%\end{eqnarray*}
%and hence, it is possible to choose $m_1(\overline{t})$ as claimed in (\ref{eq: ia-retro-coop m1eff choice dCSI+opfb 2user IC}). Thus, the first requirement of Design Criterion 3 is satisfied. The requirement of the same holds trivially.

First we show that $m_1(\overline{t}) \geq N_1$ $\forall$ $\overline{t} \in [1:t_1]$ for which it is sufficient to prove that
\[
\left\lfloor N_2 \frac{N_1-M_2}{N_2 - M_2} \right\rfloor \geq N_1.
\]
This inequality is proved as follows:
\begin{eqnarray*}
\lefteqn{ N_1 \geq N_2 \Rightarrow N_1-M_2 \geq N_2-M_2 \Rightarrow \frac{M_2}{N_2-M_2} +1 \geq \frac{M_2}{N_1-M_2} +1 } \\
&& {} \Rightarrow \frac{N_2}{N_2-M_2} \geq \frac{N_1}{N_1-M_2} \Rightarrow N_2 \frac{N_1-M_2}{N_2-M_2} \geq N_1 \Rightarrow \left\lfloor N_2 \frac{N_1-M_2}{N_2 - M_2} \right\rfloor \geq N_1,
\end{eqnarray*}
as desired. Now, making use of the fact that $m_1(\overline{t}) -N_1 \geq 0$ $\forall$ $\overline{t} \in [1:t_1]$, we obtain the following:
\begin{eqnarray*}
\Big| \mathcal{S}_{\rm LC}(b) \Big| & = & \sum_{\overline{t}=1}^{t_1} n_{\rm req}(\overline{t}) = \sum_{\overline{t}=1}^{t_1} \Big\{ m_1(\overline{t}) - N_1 \Big\} \\
& = & \sum_{\overline{t}=1}^{t_1} \Big\{ m_1(\overline{t}) \Big\} - N_1 t_1 = T d_1 - N_1 t_1 ~ \cdots \mbox{ by Design Criteria 3}\\
& = & N_2(N_1-M_2) - N_1 (N_2-M_2)   ~ \cdots \mbox{ by the choice of $t_1$ and $T$}\\
& = & M_2 (N_1-N_2) = (N_1-N_2) t_2,
\end{eqnarray*}
which proves that Design Criteria 5 is satisfied.
Since all Design Criteria 1-5 hold, we know that $P_{o2,3} \in \mathbf{D}^{\rm S}$.

\subsection{Proof of  $P_{o2,1}\in \mathbf{D}^{\rm S}$ under Case B}
Recall that under Case A, Condition $1$ holds and
\[
M_1' < N_2 \frac{N_1-M_2}{N_2 - M_2}.
\]
Here,
\[
P_{o2,1} \equiv (d_1,d_2) = \left( M_1' \frac{N_2-M_2}{N_2}, M_2 \right).
\]

We use the generic retrospective interference alignment scheme with the following choice of parameters:
\begin{eqnarray*}
T & = & N_2, ~ t_1 = N_2-M_2, ~ t_2 = M_2, \\
m_1(\overline{t}) & = & M_1' ~ \forall ~ \overline{t} \in [1:t_1], \\
m_2(\overline{t}) & = & M_2 ~ \forall ~ \overline{t} \in [1:T].
\end{eqnarray*}
It is easy to verify that Design Criteria 1-3 are satisfied. Design Criterion 4 holds because $\sum_{\overline{t}=1}^{t_1} m_2(\overline{t}) = M_2 t_1 = M_2 (N_2-M_2) = (N_2-M_2) t_2$. Consider now the proof that the last criterion is satisfied.

Here, $n_{\rm req}(\overline{t}) = M_1'-N_1$ $\forall$ $\overline{t} \in [1:t_1]$. Hence,
\[
\Big| \mathcal{S}_{\rm LC}(b) \Big| = (M_1'-N_1) t_1 = (M_1'-N_1)(N_2-M_2).
\]
Design Criterion 5 because
\begin{eqnarray*}
\lefteqn{ M_1' \leq N_2 \frac{N_1-M_2}{N_2-M_2} ~ \cdots ~ \mbox{by definition of Case B} } \\
&& {} \Rightarrow M_1' ( N_2-M_2) \leq N_2 (N_1-M_2) = (N_1-N_2) M_2 + N_1(N_2-M_2) \\
&& {} \Rightarrow (M_1'-N_1) (N_2-M_2) \leq (N_1-N_2) M_2 \\
&& {} \Rightarrow \left| \mathcal{S}_{\rm LC}(b) \right| \leq (N_1-N_2) t_2.
\end{eqnarray*}

Since all Design Criteria hold, $P_{o2,1}\in \mathbf{D}^{\rm S}$ under Case B.

\subsection{Proof of $P_{1,3} \in \mathbf{D}^{\rm S}$ under Case B}

Here,
\[
P_{1,3} \equiv (d_1,d_2) = \left( M_1' \frac{N_1-N_2}{M_1'-N_2}, N_2 \frac{M_1'-N_1}{M_1'-N_2} \right).
\]

Set
\begin{eqnarray*}
T = M_1' - N_2, ~ t_1 = N_1 - N_2, ~  t_2 = M_1'-N_1\\
\mbox{and } m_1(\overline{t}) = M_1 ~ \forall ~ \overline{t} \in [1:t_1].
\end{eqnarray*}
In order to choose $m_2(\overline{t})$, consider the following argument.
\begin{eqnarray*}
\lefteqn{ M_1' \leq N_2 \frac{N_1-M_2}{N_2-M_2} \Rightarrow M_1'(N_2-M_2) \leq N_2 (N_1-M_2) }\\
&& {} \Rightarrow N_2(M_1'-N_1) \leq M_2(M_1'-N_2) \\
&& {} \Rightarrow M_2 N_1 - M_2 M_1' + N_2(M_1'-N_1) \leq M_2 N_1 - M_2 N_2 \\
&& {} \Rightarrow M_2 (N_1-N_2) = M_2 t_1 \geq (N_2-M_2)(M_1'-N_1).
\end{eqnarray*}
Therefore, we may select $m_2(\overline{t})$, $\overline{t} \in [1:t_1]$, such that
\begin{eqnarray*}
\lefteqn{ \hspace{-4cm} 0 \leq m_2(\overline{t}) \leq M_2 ~ \forall ~ \! \overline{t} \in [1:t_1] \mbox{ and } \sum_{\overline{t} =1}^{t_1} m_2(\overline{t}) = (N_2-M_2) (M_1'-N_1); }\\
&& {} \hspace{-4cm} \mbox{and } m_2(\overline{t}) = M_2, ~ \forall \overline{t} \in [t_1+ 1:t_1+t_2].
\end{eqnarray*}
It can be easily verified that the above choice of parameters satisfies the first three design criteria. Design Criterion 4 holds because
\begin{eqnarray*}
(N_2-M_2) t_2 = (M_2-M_2) (M_1'-N_1) \sum_{\overline{t} =1}^{t_1} m_2(\overline{t}).
\end{eqnarray*}
Similarly, it is easy to verify that Design Criterion 5 holds. Therefore, we have that that $P_{1,3} \in \mathbf{D}^{\rm S}$ under Case B.

\section{Conclusion}

%We determined the DoF region of the MIMO IC with Shannon feedback.  It is shown that output feedback can improve the DoF region of the MIMO IC when there is delayed CSIT, unlike when there is instantaneous CSIT. A new IA-based achievability scheme is developed which can effectively exploit the transmitter cooperation introduced by Shannon feedback to attain the DoF gains.

In this paper, the fast fading MIMO IC is studied under the Shannon feedback setting in which the transmitters are assumed to have perfect knowledge of the channel matrices and the channel outputs, both with a finite delay. Under such a setting, the DoF region is determined with the proof involving in part the demonstration of a key achievability result that in some cases output feedback can improve the DoF region in presence of delayed CSIT. To realize the DoF gains attainable with Shannon feedback, this new achievability scheme not only employs all interference alignment techniques that are feasible with just delayed CSIT, but in addition, also exploits the additional transmitter cooperation that output feedback can induce. This result is further strengthened by identifying scenarios of limited Shannon feedback in which the entire Shannon-feedback DoF region is achievable even under the knowledge of some of the channel matrices and channel outputs at the transmitters. For example, the three DoF regions under just delayed CSIT, just output feedback, and Shannon feedback are proved to be identical for a large class of MIMO ICs. Moreover, while this work obtains the DoF region under output feedback for a large class of MIMO ICs, its complete characterization and its relationship to the delayed CSIT and Shannon feedback DoF regions remains an open problem that merits further investigation.

\appendices

\section{Proof of Lemma \ref{lem: DoF region 2user IC p-CSI and op-fb}} \label{app: proof of lem: DoF region 2user IC p-CSI and op-fb}

The region $\mathbf{D}^{\mathrm{iCSI \& op}}$ is achievable with just instantaneous CSIT \cite{Chiachi-Jafar}. Thus, it is sufficient to prove that $\mathbf{D}^{\mathrm{iCSI \& op}}$ is an outer-bound to the DoF region of the MIMO IC with instantaneous CSIT and output feedback. Toward this end, recall that the DoF achievable over the point-to-point MIMO channel can not exceed the minimum of the number of transmit and receive antennas \cite{Telatar} (this is referred as the `single-user' bound). This implies that $d_i \leq \min(M_i,N_i)$ for each $i \in\{1,2\}$. Consider now the bound on $d_1 + d_2$. Even if both transmitters and both receivers are assumed to cooperate, the total sum-DoF are limited by $M_1+M_2$ and $N_1+N_2$ due to the single-user bound. Therefore, $d_1 + d_2 \leq \min\{ M_1+M_2,N_1+N_2 \}$. Now, due to symmetry, it is sufficient to show that $d_1 + d_2 \leq \max(M_1,N_2)$. The proof of this claim, which makes use of techniques developed in \cite{Chiachi-Jafar}, is given below. %(see also Remark \ref{rem: what is different proof of lem: DoF region 2user IC p-CSI and op-fb}).

\subsection{Proof of $d_1 + d_2 \leq \max(M_1,N_2)$ with Instantaneous CSIT and Output Feedback} \label{sec: proof of lem: DoF region 2user IC p-CSI and op-fb}

As stated earlier, it is sufficient to prove that $d_1 + d_2 \leq \max(M_1,N_2)$.
%Consider the following lemma.
\begin{lemma} \label{lem: lem: DoF region 2user IC p-CSI and op-fb N_2 geq M_1}
For the MIMO IC with i.i.d. Rayleigh fading and $N_2 \geq M_1$,
\[
d_1 + d_2 \leq N_2
\]
when there is instantaneous CSIT and output feedback.
\end{lemma}
\begin{IEEEproof}
The proof of this lemma is based on the techniques developed in \cite{Chiachi-Jafar}. This lemma can not however be immediately deduced from \cite[Theorem 9]{Chiachi-Jafar} because \begin{inparaenum}[(i)] \item the model of IC with cooperation studied in \cite[Section IV]{Chiachi-Jafar} does not include the case of output feedback considered here and \item in the IC considered in \cite{Chiachi-Jafar}, the channel matrices are time-invariant and deterministic (not fading) and (iii) the IC is known to be not separable in general \cite{Cadambe_Jafar_separability, Lalitha_Poor_separability} \end{inparaenum}.
See Appendix \ref{proof of lem: lem: DoF region 2user IC p-CSI and op-fb N_2 geq M_1} for the complete proof.
\end{IEEEproof}

Thus, as per the above lemma, the required inequality holds when $N_2 \geq M_1$. When $N_2 < M_1$, as argued in \cite{Jafar-Maralle}, \cite{Chiachi-Jafar}, we may add $M_1 - N_2$ antennas at R2 (which can not reduce the DoF region), and then apply the above lemma to prove that $d_1 + d_2 \leq M_1$ if $M_1 > N_2$. Therefore, we together have $d_1 + d_2 \leq \max(M_1,N_2)$, as desired.

\subsection{Proof of Lemma \ref{lem: lem: DoF region 2user IC p-CSI and op-fb N_2 geq M_1}} \label{proof of lem: lem: DoF region 2user IC p-CSI and op-fb N_2 geq M_1}

This lemma is proved by making use of the techniques developed in \cite[Proof of Theorem 9]{Chiachi-Jafar}.

Let $U_1(t) \define H_{11}(t) X_1(t) + W_1(t)$ and $U_2(t) \define H_{21}(t) X_1(t) + W_2(t)$. Then we have the following corollary, which is stated using the notation of Section \ref{subsec: channel model Shannon journal}.
\begin{corollary}[Lemma 8, \cite{Chiachi-Jafar}]
The following is true:
\begin{eqnarray*}
\overline{X}_1(n) & \leftarrow & \mathcal{M}_1, ~ \mathcal{M}_2, ~ \overline{U}_1(n-1), ~ \overline{U}_2(n-1), ~ \overline{H}(n); \\
\overline{X}_2(n) & \leftarrow & \mathcal{M}_2, ~ \overline{U}_1(n-1), ~ \overline{U}_2(n-1), ~ \overline{H}(n); \\
\overline{Y}_1(n), ~ \overline{Y}_2(n) & \leftarrow &  \mathcal{M}_2, ~ \overline{U}_1(n), ~ \overline{U}_2(n), ~ \overline{H}(n),
\end{eqnarray*}
where $a \leftarrow b$ denotes the fact that $a$ is a deterministic function of $b$.
\end{corollary}
\begin{IEEEproof}
Can be proved via induction.
\end{IEEEproof}

We now apply Fano's inequality assuming that R1 knows the message $\mathcal{M}_2$, and signals $\overline{U}_1(n)$ and $\overline{U}_2(n)$ to obtain
\begin{eqnarray*}
n R_1 & \leq & I \Big( \mathcal{M}_1 ; \overline{Y}_1(n), \overline{U}_1(n), \overline{U}_2(n) \Big| \mathcal{M}_2, \overline{H}(n) \Big) + n \epsilon_n \\
& = & I \Big( \mathcal{M}_1 ; \overline{U}_1(n), \overline{U}_2(n) \Big| \mathcal{M}_2, \overline{H}(n) \Big) + n \epsilon_n \\
& = & h \Big( \overline{U}_1(n), \overline{U}_2(n) \Big| \mathcal{M}_2, \overline{H}(n) \Big) - h \Big( \overline{U}_1(n), \overline{U}_2(n) \Big| \mathcal{M}_2, \mathcal{M}_1, \overline{H}(n) \Big)+ n \epsilon_n,
\end{eqnarray*}
where the first equality holds since $\overline{Y}_1(n) \leftarrow \big\{ \mathcal{M}_2, \overline{U}_1(n), \overline{U}_2(n), \overline{H}(n) \big\}$. Now, following the analysis in \cite[Proof of Theorem 9]{Chiachi-Jafar}, we get
\begin{eqnarray*}
h \Big( \overline{U}_1(n), \overline{U}_2(n) \Big| \mathcal{M}_2, \overline{H}(n) \Big) & \leq & h \Big( \overline{Y}_2(n) \Big| \mathcal{M}_2, \overline{H}(n) \Big) + \sum_{t=1}^n h \Big( U_1(t) \Big| U_2(t), \overline{H}(n) \Big) \\
h \Big( U_1(t) \Big| U_2(t), \overline{H}(n) \Big) & \leq & o(\log_2 P) \\
h \Big( \overline{U}_1(n), \overline{U}_2(n) \Big| \mathcal{M}_2, \mathcal{M}_1, \overline{H}(n) \Big) & \geq & n \cdot o (\log_2 P),
\end{eqnarray*}
where $o(\log_2 P)$ is constant with $n$. These bounds give
\begin{eqnarray}
n R_1 \leq  h \Big( \overline{Y}_2(n) \Big| \mathcal{M}_2, \overline{H}(n) \Big) + n \cdot o(\log_2 P) + n \epsilon_n. \label{eq: bound on R_1 proof of lem: lem: DoF region 2user IC p-CSI and op-fb N_2 geq M_1}
\end{eqnarray}

Now Fano's inequality applied at R2 yields
\begin{eqnarray}
n R_2 & \leq & I \Big( \mathcal{M}_2 ; \overline{Y}_2(n) \Big| \overline{H}(n) \Big) + n \epsilon_n \nonumber  \\
& \leq & h \Big( \overline{Y}_2(n) \Big| \overline{H}(n) \Big) - h \Big( \overline{Y}_2(n) \Big| \mathcal{M}_2, \overline{H}(n) \Big) + n \epsilon_n \label{eq: bound on R_2 proof of lem: lem: DoF region 2user IC p-CSI and op-fb N_2 geq M_1}
\end{eqnarray}

The desired bound can now be derived by adding inequalities in (\ref{eq: bound on R_1 proof of lem: lem: DoF region 2user IC p-CSI and op-fb N_2 geq M_1}) and (\ref{eq: bound on R_2 proof of lem: lem: DoF region 2user IC p-CSI and op-fb N_2 geq M_1}), and subsequently applying the single-user bound (cf. \cite[Proof of Theorem 9]{Chiachi-Jafar}).

%\begin{remark} \label{rem: what is different proof of lem: DoF region 2user IC p-CSI and op-fb}
%While we prove Lemma \ref{lem: lem: DoF region 2user IC p-CSI and op-fb N_2 geq M_1} along the lines of the proof of Theorem 9 of \cite{Chiachi-Jafar}, Lemma \ref{lem: lem: DoF region 2user IC p-CSI and op-fb N_2 geq M_1} can not be deduced directly from \cite[Theorem 9]{Chiachi-Jafar} because of the following reasons:
%\begin{enumerate}
%\item it is not clear if the model of IC with cooperation studied in
%    \cite[Section IV]{Chiachi-Jafar} subsumes the case of output feedback considered here; and
%\item the IC considered in \cite{Chiachi-Jafar} is non-fading in the sense that the channel matrices are time-invariant and deterministic, and the IC is known to be non-separable in general \cite{Cadambe_Jafar_separability, Lalitha_Poor_separability}.
%\end{enumerate}
%Hence, Lemma \ref{lem: lem: DoF region 2user IC p-CSI and op-fb N_2 geq M_1} is not just a direct corollary of \cite[Theorem 9]{Chiachi-Jafar} and needs to be proved here explicitly.
%\end{remark}

\section{Proofs of Corollaries \ref{cor: limited and designable Shannon feedback journal} and \ref{cor: just output feedback Shannon journal}} \label{app: proof of cor: limited and designable Shannon feedback outout feedback journal}

\subsection{Proof of Corollary \ref{cor: limited and designable Shannon feedback journal}} \label{app: proof of cor: limited and designable Shannon feedback journal}

First consider the case of limited Shannon feedback. Since $ \mathbf{D}^{\rm \mathit{l}S1}, \mathbf{D}^{\rm \mathit{l}S2} \subseteq \mathbf{D}^{\rm S},$ it is sufficient to prove that the DoF regions $\mathbf{D}^{\rm \mathit{l}S1}$ and $\mathbf{D}^{\rm \mathit{l}S2}$ are achievable when there is limited Shannon feedback of Types 1 and 2, respectively. In other words, we need to prove that the region $\mathbf{D}^{\mathrm{S}}_{\mathrm{outer}}$ is achievable under two types of limited Shannon feedback, which we do next. Again, assume without loss of generality that $N_1 \geq N_2$.

If Condition $1$ does not hold, then $\mathbf{D}^{\mathrm{S}}_{\mathrm{outer}} = \mathbf{D}^{\mathrm{dCSI}}$. Then, as pointed out in \cite{Vaze_Varanasi_delay_MIMO_IC}, the region $\mathbf{D}^{\mathrm{dCSI}}$ is achievable when for each $i \in \{1,2\}$, the $i^{th}$ transmitter knows $H_{ji}(t)$ and $H_{jj}(t)$, $j \in \{1,2\}$ with $j \not= i$, with a delay. Hence, when Condition $1$ does hold, the region $\mathbf{D}^{\mathrm{S}}_{\mathrm{outer}} = \mathbf{D}^{\mathrm{dCSI}}$ is achievable with limited Shannon feedback of Types $1$ and $2$.

When Condition $1$ holds, the region $\mathbf{D}^{\mathrm{S}}_{\mathrm{outer}}$ is shown to be achievable with Shannon feedback of both types by developing a coding scheme in Section \ref{sec: proof of thm: inner-bound 2user IC d-CSI and op fb}. It can be verified that this coding scheme works even under limited Shannon feedback of both types.

Hence, together, we have proved that $\mathbf{D}^{\mathrm{S}}_{\mathrm{outer}}$ is achievable with limited Shannon feedback.

For the setting of designable Shannon feedback: Clearly, $\mathbf{D}^{\rm S} \subseteq \mathbf{D}^{\rm \mathit{d}S}$. Hence, it is sufficient to prove that the region $\mathbf{D}^{\rm S}_{\rm outer}$ is an outer-bound, even under designable Shannon feedback. Toward this end, the proof of Theorem \ref{thm: outer-bound 2user IC d-CSI and op fb} can be easily modified to apply to the general case of designable Shannon feedback.

\subsection{Proof of Corollary \ref{cor: just output feedback Shannon journal}} \label{app: proof of cor: just output feedback Shannon journal}

The first part of the corollary follows trivially since the setting of Shannon feedback is stronger than that of output feedback. To prove the second part of the corollary, we assume below without loss of generality that $N_1 \geq N_2$; and prove that the region $\mathbf{D}^{\rm S} = \mathbf{D}^{\rm S}_{\rm outer}$ is achievable with output feedback, whenever the inequality $\min(M_1,N_1) > N_2 > M_2$ does not hold (note, with $N_1 \geq N_2$, the second inequality in the statement of the corollary can never be true). This is the goal of the remainder of this sub-section.

Here, by no side-information at the transmitters, we mean the setting where the transmitters have no knowledge whatsoever of the channel states and the channel outputs; and denote the corresponding DoF region by $\mathbf{D}^{\rm no}$ which is known from the literature \cite{Vaze_Dof_final, Vaze_Varanasi_Interf_Loclzn_2011, Zhu_Guo_noCSIT_DoF_2010, Chiachi2}.

Throughput the remainder of this subsection, we assume that the inequality $\min(M_1,N_1) > N_2 > M_2$ does not hold. Then
\[
\mathbf{D}^{\rm no} \subseteq \mathbf{D}^{\rm op} \subseteq \mathbf{D}^{\rm S} = \mathbf{D}^{\rm dCSI} = \mathbf{D}^{\rm S}_{\rm outer} \subseteq \mathbf{D}^{\rm iCSI} = \mathbf{D}^{\rm iCSI\&op}.
\]

From \cite[Table I]{Vaze_Varanasi_delay_MIMO_IC}, we observe that $\mathbf{D}^{\rm no} = \mathbf{D}^{\rm dCSI}$, if the following two inequalities do not hold:
\begin{itemize}
\item $M_1 > \max(N_1, N_2)$,  $M_2 \geq N_1$, and $M_2 > N_2$ (Case A.I.3 in
        \cite[Table I]{Vaze_Varanasi_delay_MIMO_IC}); and
\item $M_1 > \max(N_1,N_2)$ and $N_1 > M_2 \geq N_2$ (Case A.II.2 in \cite[Table I]{Vaze_Varanasi_delay_MIMO_IC}).
\end{itemize}
Hence, whenever Cases A.I.3 and A.II.2 do not hold (recall the inequality $\min(M_1,N_1) > N_2 > M_2$ is not true), then
\[
\mathbf{D}^{\rm no} = \mathbf{D}^{\rm dCSI} \Rightarrow \mathbf{D}^{\rm no} = \mathbf{D}^{\rm dCSI} = \mathbf{D}^{\rm op} = \mathbf{D}^{\rm S};
\]
and thus the corollary holds.

We now proceed to Cases A.I.3 and A.II.2, under which we want to show that $\mathbf{D}^{\rm dCSI} = \mathbf{D}^{\rm op}$. Toward this end, for Case A.I.3 and A.II.2, the region $\mathbf{D}^{\rm dCSI} $ has been shown to be achievable under delayed CSIT by developing two IA-based coding scheme in \cite[Section VI]{Vaze_Varanasi_delay_MIMO_IC} and \cite[Section VII]{Vaze_Varanasi_delay_MIMO_IC}, respectively. Although these schemes have been developed there for delayed-CSIT case, they work even with output feedback. Therefore, using the schemes of  \cite[Section VI]{Vaze_Varanasi_delay_MIMO_IC} and \cite[Section VII]{Vaze_Varanasi_delay_MIMO_IC}, we conclude that $\mathbf{D}^{\rm dCSI} = \mathbf{D}^{\rm op}$ under Cases A.I.3 and A.II.2.

% *******REFERENCES*********************
\bibliographystyle{IEEEtran}
\bibliography{v6_Shannon_fb_2user_IC}
\end{document}